\documentclass[twocolumn]{bmcart}% uncomment this for twocolumn layout and comment line below
\usepackage{amsthm,amsmath,amssymb}
\usepackage[utf8]{inputenc} %unicode support

\usepackage[caption=false, font=footnotesize]{subfig}
\usepackage{booktabs}
\usepackage{flushend}

\usepackage[boldmath]{numprint}
\usepackage{stackengine}
\usepackage[acronym, nomain]{glossaries}
\usepackage{balance}
\usepackage[pdftex]{graphicx}
\usepackage{xcolor}
\usepackage{bm}
\usepackage{enumerate}
\usepackage[colorlinks=true,linkcolor=black,anchorcolor=black,citecolor=black,filecolor=black,menucolor=black,runcolor=black,urlcolor=black]{hyperref}
\usepackage{diagbox}
\usepackage{booktabs}
\usepackage{makecell}
\usepackage{tabularx,ragged2e}
\newcolumntype{?}{!{\vrule width 2\arrayrulewidth}}
\usepackage{flushend}
\usepackage[capitalise,noabbrev]{cleveref}

\def\I{\mathbf{I}}
\def\P{\mathbf{P}}
\def\K{\mathbf{K}}
\def\W{\mathbf{W}}
\def\z{\mathbf{z}}
\def\x{\mathbf{x}}
\def\T{\bm{\phi}}
\def\tpsnr{\tau_{\textrm{PSNR}}}

\newcommand{\red}[1]{\textcolor{red}{#1}}
\newcommand{\norm}[1]{\left\lVert#1\right\rVert}
\DeclareMathOperator*{\argmin}{arg\,min}
\definecolor{mat_yellow}{HTML}{ecb01f}
\definecolor{mat_purple}{HTML}{7d2e8d}
\definecolor{mat_green}{HTML}{76ab2f}
\definecolor{sec_color}{HTML}{007166}

\makeglossaries

\newacronym{prnu}{PRNU}{Photo Response Non-Uniformity}

\newacronym{spn}{SPN}{Sensor Pattern Noise}

\newacronym{ncc}{NCC}{Normalized Cross-Correlation}

\newacronym{cnn}{CNN}{Convolutional Neural Network}

\newacronym{pce}{PCE}{Peak-to-Correlation Energy}

\newacronym{psnr}{PSNR}{Peak-Signal-to-Noise-Ratio}

\newacronym{dip}{DIP}{Deep Image Prior}

\newacronym{dippas}{DIPPAS}{Deep Image Prior PRNU Anonymization Scheme}

\newacronym{roc}{ROC}{Receiver Operating Characteristic}

\newacronym{auc}{AUC}{Area Under the Curve}

\newacronym{tpr}{$\textrm{TPR}_{@0.01}$}{True Positive Rate evaluated at False Positive Rate equal to $0.01$}

\startlocaldefs
\endlocaldefs

\begin{document}

%%% Start of article front matter
\begin{frontmatter}

\begin{fmbox}
\dochead{Research}

\title{DIPPAS: A Deep Image Prior PRNU Anonymization Scheme}

\author[
   addressref={aff1},                   % id's of addresses, e.g. {aff1,aff2}
   corref={aff1},                       % id of corresponding address, if any
   %noteref={n1},                        % id's of article notes, if any
   email={francesco.picetti@polimi.it}   % email address
]{\inits{C}\fnm{Francesco} \snm{Picetti}}
\author[
   addressref={aff1},
   email={sara.mandelli@polimi.it}
]{\inits{F}\fnm{Sara} \snm{Mandelli}}
\author[
   addressref={aff1},
   email={paolo.bestagini@polimi.it}
]{\inits{P}\fnm{Paolo} \snm{Bestagini}}
\author[
   addressref={aff1},
   email={vincenzo.lipari@polimi.it}
]{\inits{A}\fnm{Vincenzo} \snm{Lipari}}
\author[
   addressref={aff1},
   email={stefano.tubaro@polimi.it}
]{\inits{S}\fnm{Stefano} \snm{Tubaro}}

\address[id=aff1]{%                           % unique id
  \orgname{Dipartimento di Elettronica, Informazione e Biongegneria - Politecnico di Milano}, % university, etc
  \street{Piazza Leonardo da Vinci, 32},                     %
  \postcode{20133}                                % post or zip code
  \city{Milano},                              % city
  \cny{IT}                                    % country
}

\begin{abstractbox}

\begin{abstract} % abstract
Source device identification is an important topic in image forensics since it allows to trace back the origin of an image.
Its forensics counter-part is source device anonymization, that is, to mask any trace on the image that can be useful for identifying the source device.
A typical trace exploited for source device identification is the Photo Response Non-Uniformity (PRNU), a noise pattern left by the device on the acquired images.
In this paper, we devise a methodology for suppressing such a trace from natural images without significant impact on image quality.
Specifically, we turn PRNU anonymization into the combination of a global optimization problem in a Deep Image Prior (DIP) framework followed by local post-processing operations.
In a nutshell, a Convolutional Neural Network (CNN) acts as generator and iteratively returns several images with attenuated PRNU traces. By exploiting a straightforward local post-processing and assembly on these images, we produce a final image that is anonymized with respect to the source PRNU, still maintaining high visual quality.
With respect to widely-adopted deep learning paradigms, the used CNN is not trained on a set of input-target pairs of images. 
Instead, it is optimized to reconstruct output images from the original image under analysis itself.
This makes the approach particularly suitable in scenarios where large heterogeneous databases are analyzed and prevents any problem due to lack of generalization.
Through numerical examples on publicly available datasets, we prove our methodology to be effective compared to state-of-the-art techniques.
\end{abstract}

%%%%%%%%%%%%%%%%%%%%%%%%%%%%%%%%%%%%%%%%%%%%%%
%%                                          %%
%% The keywords begin here                  %%
%%                                          %%
%% Put each keyword in separate \kwd{}.     %%
%%                                          %%
%%%%%%%%%%%%%%%%%%%%%%%%%%%%%%%%%%%%%%%%%%%%%%

\begin{keyword}
Deep Image Prior, Image Anonymization, Image Forensics, PRNU
\end{keyword}

% MSC classifications codes, if any
%\begin{keyword}[class=AMS]
%\kwd[Primary ]{}
%\kwd{}
%\kwd[; secondary ]{}
%\end{keyword}

\end{abstractbox}
\end{fmbox}% uncomment this for twcolumn layout

\end{frontmatter}

% \section*{Introduction}
\section{Introduction}
\label{sec:introduction}
Source device identification is a well-studied problem in the multimedia forensics community \cite{Lukas2006, Chen2008, Kirchner2019, Mandelli2020}. 
Indeed, identifying the source camera of an image helps to trace its origin and verifying its integrity.
Many state-of-the-art techniques tackle this problem by relying on \gls{prnu}, which is a unique characteristic noise pattern left by the device on each acquired content \cite{Lukas2006}. 
Given a query image and a candidate device, it is possible to infer whether the image was shot by the device with a cross-correlation test between a noise trace extracted from the image and the device \gls{prnu} \cite{Chen2007}.

Despite the effort put into developing robust \gls{prnu}-based source attribution techniques, the forensic community has also focused on studying the possibility of removing \gls{prnu} traces from images.
On one hand, determining at which level \gls{prnu} can be actually removed is essential to study the robustness of \gls{prnu}-based forensic detectors, and possibly improve them.
On the other hand, when privacy is a concern, being able to link a picture to its owner is clearly undesirable.
As an example, photojournalists carrying out legit investigations may prefer to anonymize their shots to avoid being threatened.

For these reasons, counter-forensic methods that enable deleting or reducing \gls{prnu} traces from images have been proposed in the literature.
We can broadly split the developed techniques into two families.
The first family requires the knowledge of the \gls{prnu} pattern to be deleted, and we refer to them as \gls{prnu}-aware methods.
This is the case of \cite{Karakuecuek2015, Zeng2015, Bonettini2018}, which propose different iterative solutions to delete a known \gls{prnu} from a given picture. 
Specifically, \cite{Karakuecuek2015} proposes an adaptive \gls{prnu}-based image denoising, removing an estimate of the \gls{prnu} from each image.
Authors of \cite{Zeng2015} estimate the best subtraction weight that minimizes the cross-correlation between the \gls{prnu} and the trace extracted from the image to be anonymized.
Recently, \cite{Bonettini2018} applies a \gls{cnn}, which exploits the source \gls{prnu} to hinder its traces from a query image. The network is used as a parametric operator, which iteratively overfits the given pair of image and \gls{prnu}, imposing a minimization of their correlation.

The second family of methods works by blindly modifying pixel values
%and scrambling their positions
in order to make the underlying \gls{prnu} unrecognizable.
For instance, \cite{Rosenfeld2009} shows that multiple image denoising steps can help attenuating the \gls{prnu}, even though this may not be enough to completely hinder its traces from images \cite{bernacki2020}.
Alternatively, \cite{Dirik2014} applies seam-carving to change pixel locations, and \cite{Entrieri2016} exploits patch-match techniques to scramble pixel positions.
More recently, \cite{Mandelli2017} proposes an inpainting-based method which deletes and reconstructs image pixels such that final images are anonymized with respect to their source \gls{prnu}.

In this manuscript, we propose an image anonymization tool leveraging a combination of a global optimization strategy (i.e., operating on the full image) and local post-processing operations.
In particular, global optimization on the entire image is performed exploiting a \gls{cnn}.
Given an image to be anonymized and a reference \gls{prnu} trace to be removed, the proposed network iteratively generates multiple images where \gls{prnu} traces are attenuated, still maintaining high visual quality.
Differently from most \gls{cnn} works, the proposed network does not need a training step.
In fact, the used \gls{cnn} exploits the \gls{dip} paradigm \cite{Ulyanov_2018_CVPR}, thus acting as a framework to solve an inverse problem: estimate \gls{prnu}-free images from the picture under analysis and a reference \gls{prnu} trace.
The proposed \gls{cnn} takes a random noise realization as input and iterates until it is capable of generating \gls{prnu}-free representations of a selected picture.
The analyst can decide when to stop \gls{cnn} iterations by simply checking the trade-off between the quality of the generated images and the reached anonymization level.
% accenni al local post-processing.
Then, we propose to aggregate the \gls{cnn} output images by means of a post-processing step that works at image local level. In doing so, we achieve a remarkable enhancement concerning both image quality and anonymization level, at the expense of little additional computational cost.

% In this context, image anonymization can be interpreted as an inverse problem with prior information. The prior is the network itself. Thus, the architecture plays a key role in the optimization routine.
% In order to capture the deep features from the analyzed picture, we adopt a multiresolution U-Net design \cite{ibtehaz2020multiresunet}, which has been proposed to improve the original U-Net \cite{ronneberger2015u} for multimodal medical image segmentation where the targets of interest have different shapes and scales.

The developed anonymization scheme is validated on $1200$ color images of the well known Dresden Image Database \cite{dresden} and $600$ color images of the Vision Source Identification Dataset \cite{vision}.
We address the anonymization problem on both uncompressed images (i.e., images selected from Dresden dataset) and JPEG-compressed images (i.e., images from both Dresden and Vision datasets).
For the sake of comparison with state-of-the-art techniques, we test our methodology both when an estimate of the device \gls{prnu} is available (i.e., in a \gls{prnu}-aware scenario) and when the device \gls{prnu} can be estimated only from the query image itself (i.e., in a \gls{prnu}-blind scenario).
Results show that the proposed technique actually hinders \gls{prnu}-based detectors, especially when the actual device \gls{prnu} is available.

The contributions of this paper can be summarized as follows:
\begin{itemize}
	\item We propose the first application of the \gls{dip} paradigm to a image forensic problem, to the best of our knowledge.
	\item We adapt the \gls{dip} denoising pipeline to work in case of known multiplicative noise.
	\item We propose a \gls{prnu}-aware image anonymization technique that outperforms the state-of-the-art on the Dresden dataset and is the runner up on Vision.
	\item We propose a \gls{prnu}-blind image anonymization technique that reduces \gls{prnu} traces around image edges in contrast to other methods in the literature.
	\item We propose a method that allows forensic analysts to select the trade-off between image quality and anonymization capability depending on the working scenario.
\end{itemize}

The rest of the paper is organized as follows.
In \cref{sec:background}, we provide the reader with the background concepts useful for understanding the core of the proposed methodology.
In \cref{sec:methodology}, we present the details of the proposed scheme. In particular, we first define the inverse problem we aim at solving, then we devise a processing pipeline to obtain the target \gls{prnu}-free image. Finally, we describe the architecture design along with the optimization strategies.
In \cref{sec:experiments}, we describe the experimental setup; in \cref{sec:results}, we discuss all the achieved results, compared with State of the Art solutions.
Eventually, in \cref{sec:conclusions} we draw our conclusions.

\section{Background and problem statement}
\label{sec:background}
\glsresetall

In this section, we introduce some background concepts
useful to understand the rest of the paper. 
First, we introduce \gls{prnu} and its use for in source device identification. Then, we present the adopted methodology known as \gls{dip} \cite{Ulyanov_2018_CVPR}, which has been recently proposed as a paradigm to solve diverse inverse problems like image denoising.
Eventually, we provide the formulation of the source device anonymization problem faced in this paper.
%In this section we devise the adopted methodology known as \gls{dip} \cite{Ulyanov_2018_CVPR}, which has been proposed as a paradigm to solve inverse problems, and we specialize it for the goal of PRNU attenuation, interpreted as an image restoration task.

\subsection{Photo Response Non-Uniformity}
\glsreset{prnu}
\gls{prnu} is a characteristic noise fingerprint introduced in all images acquired by a device.
%due to some imperfections in the sensor manufacturing process.
Specifically, the \gls{prnu} $\K$ has the form of a zero-mean pixel-wise multiplicative noise. 
According to the well-known model proposed in \cite{Lukas2006, Chen2008}, a generic image $\I$ shot by a digital device can be described as
% Considering a device, let $\K$ be its PRNU pattern and $\I_0$ a captured image. We can state that: 
\begin{equation}
	\I = \I_0 \left( 1 + \gamma \K \right) + \boldsymbol{\Theta}
	\label{eq:I_prnu_def}
\end{equation}
where $\I_0$ is the sensor output in the absence of noise, $\gamma$ is the weight of the \gls{prnu} contribution, and $\boldsymbol{\Theta}$ includes all the additional independent random noise components.
%where $\I$ is the PRNU-free image and $\gamma$ is the weight of the PRNU contribution.
The \gls{prnu} $\K$ can be estimated by collecting a set of images shot by the device, following the method proposed in \cite{Lukas2006, Chen2008}.  

\gls{prnu} $\K$ is commonly used to solve source device identification problem, that is, given a query image $\I$ and a candidate device, understanding if the device shot that image or not.
One way to solve the problem is to compute the \gls{ncc} \cite{Lukas2006} between a noise residual $\W$ \cite{Chen2008} extracted from the query image $\I$ and the \gls{prnu} $\K$ of the candidate device, pixel-wise scaled by $\I$. 
Referring to \cite{Lukas2006}, we can define the \gls{ncc} as
\begin{equation}
	% \text{NCC}(\W, \I \K) = \frac{\mathrm{corr}([\W]_{r,c}, [\I \K]_{r,c})}{\Vert \W \Vert_{\mathrm{F}} \cdot \Vert \I \K \Vert_{\mathrm{F}}}
	% \text{NCC}(\W, \I \K) = \frac{\mathrm{corr}(\W, \I \K)}{\Vert \W \Vert_{\mathrm{F}} \cdot \Vert \I \K \Vert_{\mathrm{F}}},
	\text{NCC}(\W, \I \K) = \frac{\sum_{i,j} [\W]_{i,j} \cdot [\I \K]_{i,j}}{\Vert \W \Vert_{\mathrm{F}} \cdot \Vert \I \K \Vert_{\mathrm{F}}},
	\label{eq:ncc_def}
\end{equation}
where $|| \cdot ||_{\mathrm{F}}$ is the Frobenius norm, while $[\W]_{i,j}$ and $[\I \K]_{i,j}$ are the terms in position $(i , j)$ of $\W$ and $\I \K$, respectively.
If $\text{NCC}(\W, \I \K)$ is greater than a predefined threshold, the image can be attributed to the device with a certain confidence.

\subsection{Deep Image Prior}
\glsreset{dip}
%The 
A generic image restoration problem is
%An inverse problem is
usually solved through the minimization of an objective function of the form
\begin{equation}
	J(\x)=E(\x;\I)+\lambda R(\x),
	\label{eq:obj}
\end{equation}
where $E(\x;\I)$ is a task-dependent misfit function conditioned by the (corrupted) 
input image $\I$ and $R(\x)$ is a regularization term designed to tackle the ill-posedness and ill-conditioning of the inverse problem. To avoid confusion with the rest of notation used in the paper, we refer at $\x$ as a generic image which the cost function is evaluated for. 
$\lambda$ is a weight setting the trade-off between honoring the data and imposing the desired a-priori features \cite{moulin1999analysis}.
The restored image $\hat{\x}$ is then obtained as
\begin{equation}
	\hat{\x} = \argmin_{\x} J(\x).
	\label{eq:min_obj_reg}
\end{equation}

The data misfit $E(\x;\I)$ is usually quite simple to devise, and it depends on the desired task.
The design of a regularization term $R(\x)$ can be challenging because it should capture the features of the desired image.
% In the past years, the imaging community has focused its efforts on designing regularizers able to enforce a-priori knowledge on the 
% image
% %model
% to be retrieved.
% Among them, Tikhonov regularization \cite{tikhonov1977solution} privileges minimum energy solutions and Total Variation (TV) \cite{rudin1992nonlinear} privileges flat-zones solutions.
% Lately, more elaborated regularizers have been based on the idea that the image can be conveniently represented in a transformed domain such as the Fourier domain, or using Wavelet-based representations \cite{malfait1997wavelet}, Curvelets \cite{starck2002curvelet}, Contourlets \cite{do2005contourlet}, and Shearlets \cite{guo2007optimally}.

\gls{dip} has been proposed as an alternative solution with respect to standard regularization \cite{Ulyanov_2018_CVPR}.
The objective function to minimize is recast as
\begin{equation}
	J(\T)=E(\x;\I)=E(f_{\T}(\z);\I),
	\label{eq:dip}
\end{equation}
where $f_{\T}(\cdot)$ is a \gls{cnn} represented as a parametric non linear function, $\T$ are the parameters of this function (i.e., the weights of the \gls{cnn}), and $\z$ is a random noise realization. The value $f_{\T}(\z)$ is associated with the output image to the network, thus it is related to a random noise realization $\z$ and to the \gls{cnn} parameters $\T$.
%In a DIP-based framework, the minimization is performed in the space of cnn parameters $\T$.
Notice that there is not an explicit regularization term: the \gls{cnn} architecture itself plays the role of the prior. Through its convolutional layers, the \gls{cnn} captures the inner structure and self-similarities of the desired uncorrupted image from the input corrupted one, and constraints the solution space.
In other words, instead of minimizing the objective function in the space of the image as in \eqref{eq:min_obj_reg}, \gls{dip} performs the search in the space of the \gls{cnn} parameters $\T$. This dramatically changes the shape of the objective function, driving the solution to honor both the data misfit and the deep features captured on the corrupted input image. 
The restored image is then obtained as
\begin{equation}
	\hat{\x}=f_{\hat{\T}}(\z),
	\label{eq:output}
\end{equation}
where
\begin{equation}
	\hat{\T}= \argmin_{\T} J(\T).
\end{equation}

% It is worth noticing that the \gls{cnn} optimization is not performed aiming at reconstructing a target image starting from a ground truth image.
% Indeed, the \gls{dip} method does not require a specifically designed set of data for training.
% Even though the result is the output of a \gls{cnn}, \gls{dip} does not follow a typical deep learning paradigm.

%\subsection{DIP for denoising}
For the specific case of image denoising, the DIP objective function presented in \eqref{eq:dip} is customary set to the $\ell_2$ distance between the output image to the \gls{cnn} for a given combination of noise realization $\z$ and network parameters $\T$, defined as $f_{\T}(\z)$, and the 
%noisy
input
image $\I$ \cite{tirer2020back, dong2019denoising, lin2020hyperspectral}. For the sake of notation, from now on we refer to the \gls{cnn} output image $f_{\T}(\z)$ as $\I_{\T}$. Therefore, \eqref{eq:dip} becomes
\begin{equation}
	J(\T) = \norm{\I_{\T} -\I}_2^2.
	\label{eq:dip_denoise}
\end{equation}

One may ask why a \gls{cnn} that is designed to reconstruct a generic image should perform denoising while its goal is set to fit the
%noisy 
input (noisy)
image $\I$.
The main reason is the different behaviour of signal and noise components throughout the iterative optimization \cite{aggarwal2019modl}. If the minimization is led to convergence, the result will indeed fit the noisy image, but the authors of \cite{Ulyanov_2018_CVPR} have shown that parametrizing the optimization via the weights of a \gls{cnn} generator distorts the search space so that in the minimization process the signal fits faster than the noise.
Therefore, \cite{Ulyanov_2018_CVPR} proposes to perform denoising by early stopping the iterative minimization.
For example, in a forensic scenario, the analyst can stop the optimization when some task-specific average metrics reach a desirable value.

\subsection{Problem formulation}
\label{subsec:problem}
In this paper, we focus on the forensics counter-part of the source device identification problem, that is, performing source device anonymization. 
Specifically, given an image, we aim at hindering \gls{prnu} traces left on the image in order to make it impossible to associate the image with its original source device. Meanwhile, the visual quality of the anonymized image should not be compromised by the anonymization process.
This translates into fulfilling two main goals:
\begin{enumerate}
	\item the \gls{ncc} between the anonymized image and the actual source PRNU should be lower than a predefined threshold;
	\item the \gls{psnr} between the anonymized image and the original image should assume high values.
\end{enumerate}
To this purpose, we propose the \gls{dippas}, which is an anonymization method based on a \gls{dip} optimization framework followed by a local post-processing and assembly operations studied on purpose.
In the next section, we present the proposed strategy, discussing the
main intuitions behind the approach.

\section{Proposed Methodology}
\label{sec:methodology}

The proposed method for image anonymization works in two main steps: a \gls{dip}-based method is adapted to generated multiple images with limited \gls{prnu} traces; these images are combined into a single one by means of a post-processing scheme.

To illustrate all the details of the proposed approach, in this section, we start showing the theoretical \gls{dip}-based framework chosen for our specific problem.
Then, we explain how it is possible to generate an image using this framework.
%As \gls{dip} can return multiple images where \gls{prnu} traces are attenuated,
We then report the details of the developed post-processing scheme that merge multiple images into one.
%that makes the most out of this feature to strengthen the anonymization capability.
Finally, we describe the employed \gls{cnn} architecture.

\subsection{DIP-based image generation}
\label{subsec:dip_anonymization}

Considering the model \eqref{eq:I_prnu_def} of a generic image $\I$ acquired by a digital device, the anonymization task consists in estimating the ideal \gls{prnu}-free image $\I_0$. Indeed, $\I_0$ is completely uncorrelated from the device \gls{prnu}, and it has a reasonably good visual quality.
We aim at this goal by combining the \gls{dip} denoising paradigm of~\eqref{eq:dip_denoise} with the \gls{prnu}-based image modeling proposed in~\eqref{eq:I_prnu_def}.
$\I_{\T} = f_{\T}(\z)$ being the output of the \gls{cnn} for a given parameter configuration $\T$ and $\z$ a noise realization, the functional to be minimized becomes
\begin{equation}
	J(\T) = \norm{\I_{\T} \left( 1+\gamma \P \right) -\I}_\mathrm{F}^2.
	\label{eq:dip_prnu}
\end{equation}
We define $\P$ as the device fingerprint that can be either the estimated \gls{prnu} pattern (i.e., $\P=\K$), or the noise residual $\W$ extracted from $\I$ as suggested in \cite{Chen2008}.

%\subsubsection{PRNU-aware Anonymization}
The former situation is a \gls{prnu}-aware scenario (e.g., a user wants to anonymize its own pictures and knows the reference \gls{prnu}).
In this case, the proposed scheme makes use of the \gls{prnu} $\K$ as the fingerprint $\P$, hence \eqref{eq:dip_prnu} becomes
\begin{equation}
	J(\T) = \norm{\I_{\T} \left( 1+\gamma \K \right) -\I}_\mathrm{F}^2.
\end{equation}

%\subsubsection{PRNU-blind Anonymization}
The latter situation is a \gls{prnu}-blind scenario (e.g., a website wants to store anonymized images uploaded by users but each reference \gls{prnu} is not known at server side).
In this case, the fingerprint $\P$ we inject in the inverse problem is the noise residual $\W$ extracted from $\I$ itself \cite{Chen2008}, hence \eqref{eq:dip_prnu} becomes
\begin{equation}
	J(\T) = \norm{\I_{\T} \left( 1+\gamma \W \right) -\I}_\mathrm{F}^2.
	\label{eq:dip_prnu_blind}
\end{equation}

Notice that the term $\I_{\T} \left( 1+\gamma \P \right)$ emulates the image modeling shown in~\eqref{eq:I_prnu_def} (correctly if $\P = \K$, approximately if $\P = \W$). 
The more $\I_{\T} \left( 1+\gamma \P \right)$ approaches $\I$ in terms of Frobenius norm, the more $\I_{\T}$ will represent a reasonably better estimate of the ideal PRNU-free image $\I_0$, apart from independent random noise contributions.
Given these premises, the estimated image $\I_{\T}$ is a good candidate for the anonymization of the input image $\I$.

\begin{figure}[t]
	\centering
	\includegraphics[width=.95\columnwidth]{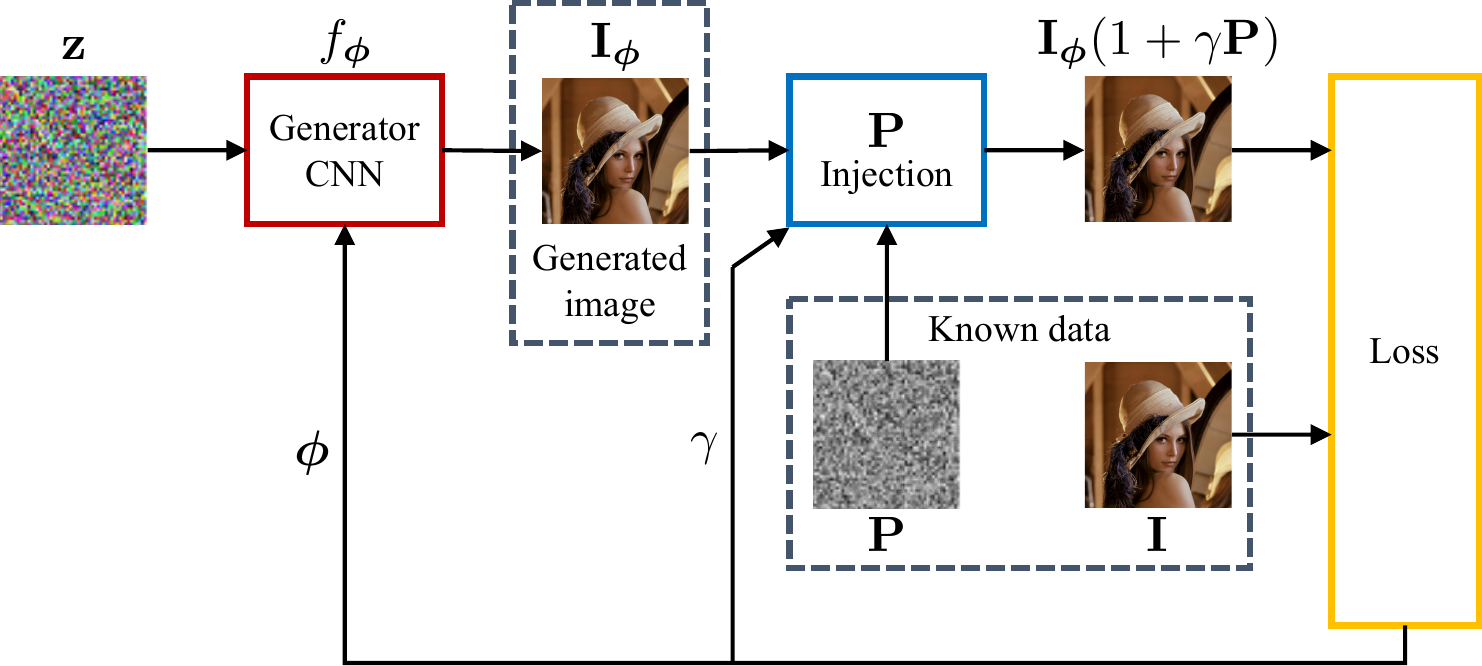}
	\caption{The proposed \gls{dip}-based image generation process. The known data are the acquired image $\I$ and the device fingerprint $\P$. During the inversion, the traces of $\P$ are attenuated by injecting $\P$ into the image $\I_{\T}$ generated by the \gls{cnn}. In the \gls{prnu}-aware setup, $\P=\K$; in the \gls{prnu}-blind setup, $\P=\W$, a noise residual extracted from $\I$ \cite{Chen2008}.}
	\label{fig:dip_scheme}
\end{figure}

% spiegazione dell'idea:
In a nutshell, the proposed strategy is depicted in \cref{fig:dip_scheme}. 
Starting from image $\I$, we extract the device \red{noise residual} $\P$ either in \gls{prnu}-aware or \gls{prnu}-blind scenario. 
Then, we generate the image $\I_{\T}$ following the \gls{dip} paradigm, i.e., imposing the fingerprint-injected image $\I_{\T} \left( 1+\gamma \P \right)$ to be as similar as possible to the known image $\I$.
By minimizing the functional \eqref{eq:dip_prnu}, we estimate the image $\I_{\T}$ with attenuated fingerprint traces.

\subsection{Generation of PRNU-attenuated images}

% specificare che possiamo generare più immagini
As we previously reported, the proposed \gls{dip} process can generate multiple images with attenuated \gls{prnu} traces.
Referring to \cref{fig:dip_scheme}, the generation pipeline is the following:
\begin{enumerate}
	% \item We normalize the input query image $\I$ in the range $[0,1]$ to better adapt to the \gls{cnn} computation dynamics.
	\item The \gls{cnn} input tensor $\mathbf{z}$ is a realization of a zero-mean white gaussian noise,
	%with zero mean and standard deviation $0.1$,
	with the same size of image $\I$. We found uniform distributions to be less effective; we are convinced the white gaussian noise is able to excite a broader frequency range and can produce better images.
	%Additionally, at each iteration we perturb the CNN input noise with additional gaussian noise with standard deviation of 0.1 to strengthen the convergence.
	\item We optimize the weights of the \gls{cnn} %through the ADAM algorithm
	by minimizing~\eqref{eq:dip_prnu}. The optimization is performed over the generator weights $\T$. Notice that the PRNU injection weight $\gamma$ acts as a trainable layer of the architecture, so $\gamma$ is estimated directly during the inversion. Specifically,  $\gamma$ is clamped to be positive, as negative $\gamma$ values are not model representative.
	%in order to improve the interpretability.
	%The parameter $\gamma$ can be either set by the analyst or estimated by a specifically designed layer of the network itself. Indeed, the modeling can be recast as the weighted Hadamard product between $\P$ and $\I_{\T}$.
	\item At each minimization step, we generate an image $\I_{\T}$, which 
	%the generated image $\I_{\T}$ 
	is saved only if the \gls{psnr} with respect to the original image $\I$ is above a certain threshold $\tpsnr$. This is done to guarantee a sufficiently good visual quality for the generated image.
\end{enumerate}
\begin{figure}[t]
	\centering	\includegraphics[width=.95\columnwidth]{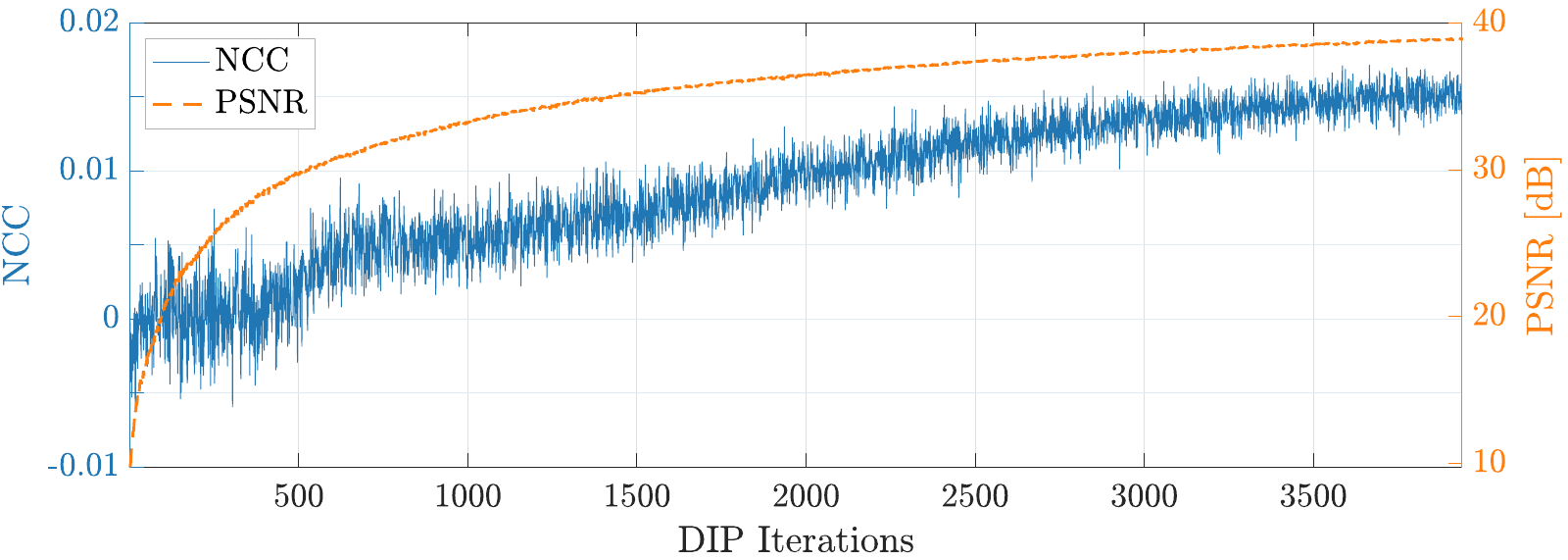}
	\caption{\gls{ncc} and \gls{psnr} behaviour of generated images as a function of \gls{dip} iterations. Some generated images are depicted in \cref{fig:dip_inversion_example_img}.}
	\label{fig:dip_inversion_example}
\end{figure}

\begin{figure*}[t]
	\centering	\includegraphics[width=.95\textwidth]{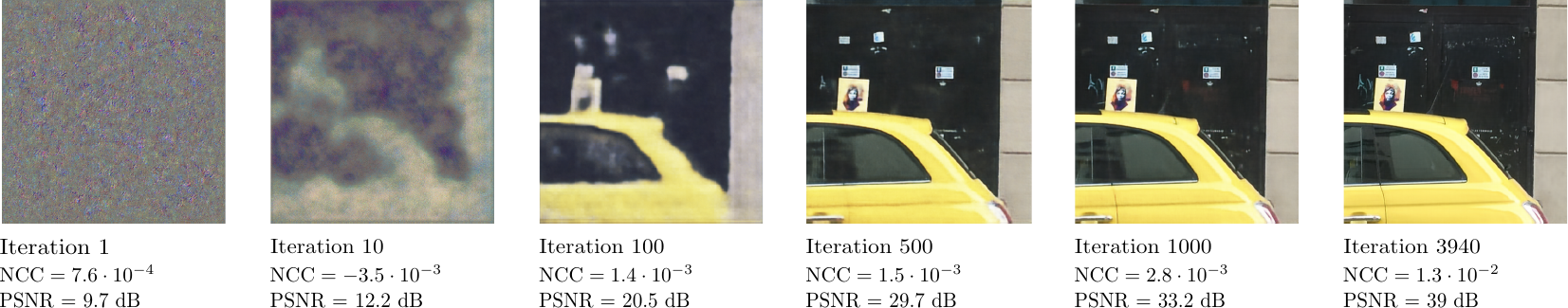}
	\caption{\gls{dip} inversion example: as iterations increase, the reconstructed images pass from a noisy behaviour (i.e., iteration $1$) to a very similar copy of the original image. However, as \gls{psnr}, \gls{ncc} can slightly grow as well.}
	\label{fig:dip_inversion_example_img}
\end{figure*}

The minimization process ends when the \gls{psnr} between the generated image $\I_{\T}$ and the initial image $\I$ overcomes a threshold of $39 \textrm{dB}$. The maximum number of iterations is anyway fixed to $10000$.
In doing so, after the \gls{dip} process ends, a pool of $M$ generated images ${\I_{\T}}^{(m)}, m \in [1, M]$ with \gls{psnr} $\geq \tpsnr$ has been collected.
For the sake of clarity, \cref{fig:dip_inversion_example} and \cref{fig:dip_inversion_example_img} report one example of the inversion process, showing the evolution of the \gls{cnn}-generated images, together with their \glspl{psnr} with respect to the original image and \glspl{ncc} with the source device \gls{prnu}.
%as a function of iterations, respectively.

% COMMENTO DA TENERE PER EVENTUALI REVIEWS
%Because the PSNR is related to the cost function \eqref{eq:dip_prnu}, it
%\red{Notice that PSNR reaches values above $30\textrm{dB}$ as the inversion goes on; in fact, the images produced after $500$ iterations are visually close to the reference one. However, as reported in \cref{fig:dip_inversion_example}, also the NCC with the device PRNU increases.
%Our interpretation is that injecting the \red{noise residual} $\P$ makes the objective function to decrease as the produced image $\I_{\T}$ will show also the noise patterns present in the original image $\I$; therefore, the NCC will increase because of the optimization process itself.}

%%%%%%%%%%%%%% al massimo solo questa parte:
% It is worth noticing that, although the proposed approach involves the optimization of \gls{cnn} parameters, it is not trained to reconstruct a target image by learning from ground truth samples.
% The deep features are learnt out of the input image $\I$ by minimizing the functional $J(\T)$ in a way that can be seen as overfitting. 
% Contrarily to standard deep learning paradigms, the optimization is performed in the space of the \gls{cnn} parameters instead of the image space.
%%%%%%%%%%%%%%

%However, \gls{dip} belongs to the context of inverse problems, where of course the modeling fits the acquired image.
%The main difference is that the optimization is performed in the space of the \gls{cnn} parameters instead of the image space.

\begin{figure*}[t]
	\centering	\includegraphics[width=.95\textwidth]{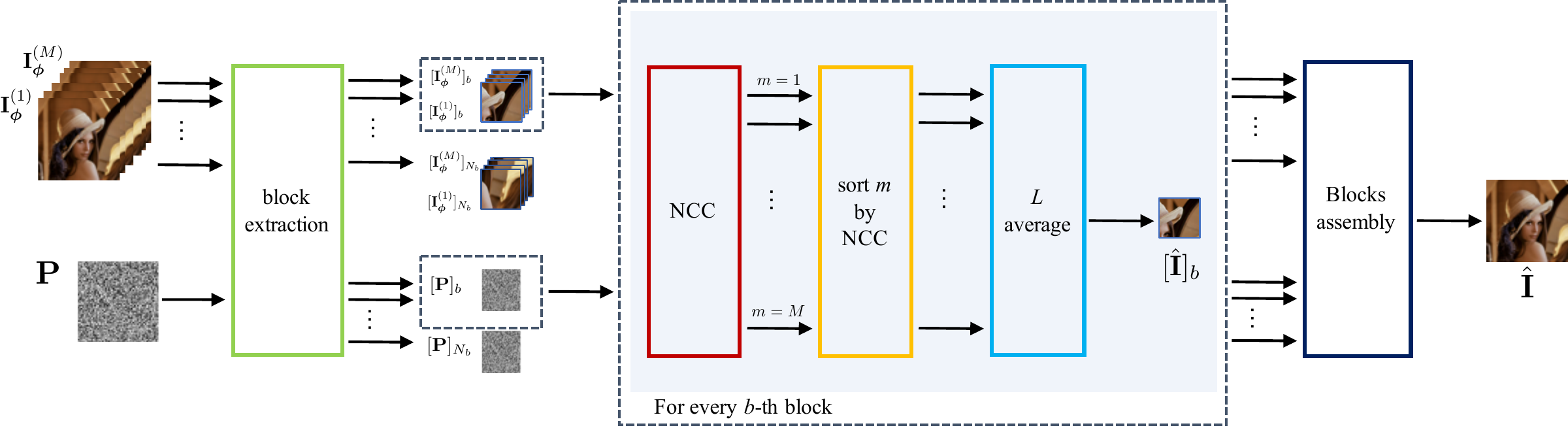}
	\caption{Assembly of the anonymized image. For each generated image ${\I_{\T}}^{(m)}$, $N_b$ blocks are extracted from the image and the fingerprint $\P$. Fixing a block position $b$, we compute the NCC between each pair of blocks $[{\I_{\T}}^{(m)}]_b$, $[\P]_b$, $m \in [1, M]$. Then, we order the $M$ image blocks according to their resulting NCCs and we average the first $L$ blocks pixel by pixel, obtaining the estimated block $[{\hat{\I}}]_b$. We follow this pipeline for each block position $b$, eventually assembling the results and estimating the anonymized image ${\hat{\I}}$.}
	\label{fig:anon_scheme_esteso}
\end{figure*}

\subsection{Local post-processing and assembly}
\label{subsec:assembly}

Notice that the \gls{dip} minimization functional \cref{eq:dip_prnu} imposes a constraint on the Frobenius norm of the difference between the fingerprint-injected image and the initial image.
This constraint represents a global constraint as it does not specifically focus on local pixel areas. 
Recalling our final goals (i.e., maximizing the PSNR and minimizing the \gls{ncc} of the anonymized image), 
%reported in \cref{subsec:problem}, which consist in (i)minimizing the NCC of the anonymized image; (ii) maximizing the PSNR of the anonymized image.
while the \gls{psnr} is a global metrics as it considers the entire image and not local areas, the \gls{ncc} can strongly depend on specific local regions of the image which can correlate with the device \gls{prnu} in diverse fashions.  

% moreover, citando le figure 2-3, possiamo notare come all'aumentare delle iterazioni, anche la NCC possa crescere
Moreover, as \gls{dip} iterations increase, the minimization process risks to inject an excessive amount of \gls{prnu} traces in the generated images.
This is noticeable from the examples depicted in \cref{fig:dip_inversion_example} and \cref{fig:dip_inversion_example_img}: when iterations' number grows, the \gls{psnr} of the generated images improves but their \gls{ncc} can slightly grow as well, resulting in worse anonymization.

Given these premises, we propose to further optimize our solution by investigating the $M$ generated images on their local areas. 
%Indeed, all the $M$ available images represent reasonable and valuable solutions for the anonymization.
We improve upon these results with
%To clarify, we propose
a very straightforward methodology to generate one final anonymized image out of the $M$ previously generated, by locally optimizing the cross-correlation with the reference device fingerprint $\P$. 
%Given these premises, we propose a very simple methodology to generate the final anonymized image $\hat{\I}$. 

Specifically, \cref{fig:anon_scheme_esteso} depicts the proposed pipeline: we divide each available image ${\I_{\T}}^{(m)}$ and the reference fingerprint $\mathbf{P}$ into $N_b$ non overlapping squared blocks of $B \times B$ pixels. Image and fingerprint blocks are defined as $[{\I_{\T}}^{(m)}]_b$ and $[\P]_b, b \in [1, N_b]$, respectively. 
Notice that, for each block geometric position $b$, we have $M$ available image blocks associated with the $M$ image realizations produced during the \gls{dip} iterations.
For each block position $b$, three main steps follow:
\begin{enumerate}
	\item We compute the \glspl{ncc} between the $M$ image blocks $[{\I_{\T}}^{(m)}]_b, m \in [1, M]$ and the fingerprint block $[\P]_b$ as in \eqref{eq:ncc_def}. 
	\item %The resulting \glspl{ncc} are analyzed as a function of the minimization step $m$. Precisely, 
	The available $M$ blocks are ordered accordingly to their resulting \glspl{ncc}: first, we select the blocks with negative \gls{ncc} and increasing absolute value; secondly, we select the blocks with positive \gls{ncc} and increasing absolute value. In doing so, blocks with low absolute value of NCC and negative NCC are given higher priority than blocks returning bigger NCCs.
	\item Following the order specified above, we average the first $L$ blocks pixel by pixel, ending up with a $B \times B$ final reconstructed block.
\end{enumerate}
The final anonymized image $\hat{\I}$ is estimated by assembling the results obtained for each single block position $b$ and color channel.
%\begin{algorithm}
%\caption{Anonymization scheme}
%\begin{algorithmic}[]
%\State $k=1$
%\State set PSNR threshold $\tau$
%\Procedure{DIP inversion}{$\I_0,\P$}
%    \For{$k=0,k < K, k++$}
%        \State $\I=f_{\T}(\z)$
%        \If{$\text{PSNR}(\I, \P)>\tau$}
%         \State extract $B\times B$ blocks from $\I, \P$
%            \State store $\text{NCC}^{(k)}(\I,\P) \; \forall$  blocks
%        \EndIf
%        \State update $\T$ \Comment{ADAM optimization}
%    \EndFor
%\EndProcedure
%\Procedure{assembly}{NCC, blocks}
%    \For{each block position $b$}
%        \State sort $b^{(k)}$ for increasing $\vert \text{NCC}^{(k)}\vert$ (negative first)
%        \State $b_A \leftarrow \frac{1}{N}\sum_n b^{(n)}$
%    \EndFor
%\EndProcedure
%	
%	\end{algorithmic}
%	\label{alg:anonym_scheme}
%\end{algorithm}
\subsection{CNN Architecture}
The U-Net is a convolutional autoencoder (i.e., a CNN aiming at reconstructing a processed version of its input) characterized by the so called skip-connections and originally introduced for medical image processing \cite{ronneberger2015u}. If properly trained according to the standard deep learning paradigm, it proves very effective for multidimensional signal processing tasks such as denoising \cite{Zhang2017, cruz2018nonlocality}, interpolation \cite{kong2020deep, mandelli2018seismic}, segmentation \cite{zhou2020unet++, jurdi2020bbunet}, inpainting \cite{sidorov2019deep}, and domain-specific post-processing operators \cite{picetti2019seismic, devoti2019wavefield}.

More recently, the Multi-Resolution U-Net \cite{ibtehaz2020multiresunet} has been proposed for multimodal medical image segmentation, based on the consideration that the targets of interest have different shapes and scales.
If we want to capture self-similarities of natural images to be employed as a prior, working at different scales can be strongly beneficial. Preliminary experiments on a small dataset led us to adopt such architecture rather than traditional U-Net and its derivations analyzed in \cite{Ulyanov_2018_CVPR}.

\begin{figure}[t]
	\centering	\includegraphics[width=.95\columnwidth]{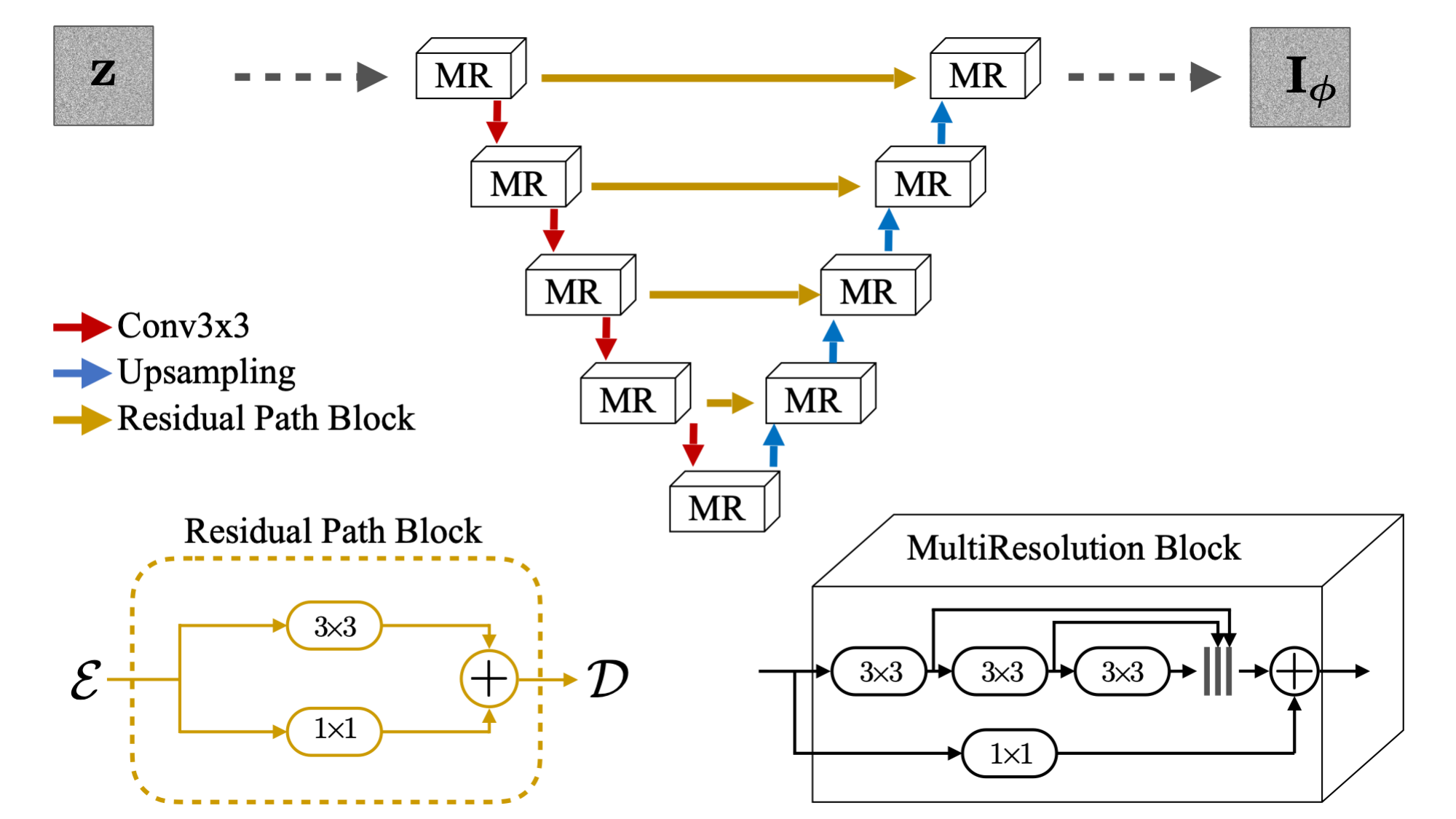}
	\caption{Proposed Multi-Resolution U-Net architecture.}
	\label{fig:multires}
\end{figure}

Therefore, we propose an ad-hoc Multi-Resolution U-Net (shown in \cref{fig:multires}) that can be summarized as follows:
\begin{enumerate}
	\item Convolutional layers are replaced by so called Multi Resolution blocks shown in the bottom-right portion of \cref{fig:multires}. These blocks approximate multi-scale features of the Inception block \red{\cite{inception}} while limiting the number of parameters of the network, which is critical when employing it as a deep prior. Every block is a chain of three convolutions; the final output is the sum of the input, scaled by a learned factor, and the stack of the three partial outputs. 
	\item Skip connections, which are the distinctive feature of the U-Net, are replaced by Residual Path blocks shown in the bottom-left portion of \cref{fig:multires}, \red{as proposed in \cite{ibtehaz2020multiresunet}}. $\mathcal{E}$ is the output of an encoding layer and $\mathcal{D}$ is concatenated to the corresponding decoding layer.
	\item Downsampling is achieved by $3\times3$ convolutions with stride $2\times2$. Upsampling is performed by nearest neighbor interpolation. Batch Normalization and LeakyReLU activation follow every convolution apart from the last one (i.e., the \gls{cnn} output) that is activated by a sigmoid.
\end{enumerate}

Notice that, even though the result is the output of a \gls{cnn}, the \gls{dip} method does not exploit the typical deep learning paradigm where a training phase is performed over a specifically designed set of data.
In particular, only the query image is used in the reconstruction process and the \gls{cnn} implicitly assumes the role of prior information that exploits correlations in the image to learn its inner structure.
% qui ci tiriamo la zappa sui piedi. O che la motiviamo meglio, o togliamo l'ultima frase.
%Therefore, the choice of a specific \gls{cnn} architecture is crucial for a suitable and well-performing solution.

\section{Experiments}
\label{sec:experiments}

In this section, we describe the used datasets, the experimental setup and the evaluation metrics.

\subsection{Datasets}
We resort to two well-known datasets, commonly used for investigating PRNU-related problems on images.
The first dataset is the Dresden Image Database \cite{dresden}, which collects both uncompressed and compressed images from more than $50$ diverse devices.
Following the same procedure done in past works proposed in literature \cite{Mandelli2017, Entrieri2016}, we select images from $6$ different camera instances, precisely Nikon D70, Nikon D70s, Nikon D200, two devices each.
Second dataset is the recently released Vision Dataset \cite{vision}, which includes JPEG compressed images captured from $35$ devices. Among the pool of available models, we collect images from $6$ different camera vendors, precisely from devices named as D12, D17, D19, D21, D24, D27 in \cite{vision}.

The PRNU fingerprint of each device is computed by collecting all the available flat-field images shot by the device and following the Maximum Likelihood estimation proposed in \cite{Chen2008}. Concerning the Dresden dataset, we exploit never-compressed Adobe Lightroom images to compute the PRNU, as it reasonably is the most accurate way to estimate the device fingerprint. Indeed, JPEG compression can create blockiness artifacts that may hinder PRNU estimation \cite{Chen2008}.
% from chen2008:
% We note that strong JPEG compression also creates blockiness artifacts that can propagate into the estimated PRNU factor
Every device includes $25$ homogeneously lit flat-field images for the PRNU estimation. 
For Vision dataset, we have more than $95$ JPEG flat-content images to compute each device fingerprint.

The images to be anonymized are selected from natural images, precisely we pick $100$ natural images per device.
Regarding Dresden dataset, two different sub-sets can be extracted: for every device, we select $100$ never-compressed Adobe Lightroom images, together with other $100$ taken from the pool of JPEG compressed images.
We end up with three distinct datasets comprising $600$ images each: the Dresden uncompressed dataset, called $\mathcal{D}_u$; the Dresden compressed dataset, defined as $\mathcal{D}_c$; the Vision (compressed) dataset, $\mathcal{V}$.

\subsection{State-of-the-arst solutions}
As state-of-the-art solutions, we select the most recent anonymization methods proposed in the literature.

% PRNU-aware
Among the pool of PRNU-aware methods, we implement the method proposed in \cite{Karakuecuek2015}, being the most recent and cited contribution.
%\red{@Paolo: verifica commenti su Nico, non vorrei tirargli la zappa sui piedi}
We do not compare our solution with the PRNU-aware strategy recently proposed in \cite{Bonettini2018}, as its performance drops significantly whenever the used image denoising operator during cross-correlation tests is the commonly used one suggested in \cite{Lukas2006, Chen2008}.
Since in our proposed strategy we follow the methodology devised in \cite{Lukas2006, Chen2008} for image denoising and cross-correlation, a comparison with \cite{Bonettini2018} would be unfair.

% PRNU-blind
Regarding PRNU-blind strategies, the most recent contribution is that proposed by \red{us in} \cite{Mandelli2017}, which demonstrates to outperform results of \cite{Entrieri2016} in a PRNU-blind scenario. For the implementation of \cite{Mandelli2017}, we consider the parameter configurations achieving the best anonymization results, i.e., the strategies defined as $\ell^{(3)}_1$ and $\ell^{(5)}_1$ in the original paper.

Moreover, to show that simple denoising does not achieve good anonymization performances \cite{bernacki2020}, we implement the well-known DnCNN denoiser \cite{Zhang2017} which represents a modern data-driven solution among image denoising strategies.  

\subsection{Experimental Setup}

% An interesting aspect of the \gls{dip} paradigm is that the \gls{cnn} can process images of any size. Obviously, the bigger the input image, the heavier the computation.
% Indeed the computing node has to store in memory all the \gls{cnn} weights.
% However, it is worth considering that the proposed method can also be applied in a patch-wise fashion and its parallelization is straightforward.
% Given these premises and c
Considering what was done in the past state-of-the-art \cite{Mandelli2017}, our experiments process images of $512\times512$ pixels with $512$ features extracted at the first MultiRes block.
To do so, we center-crop all the images and the computed PRNUs to a common resolution of $512 \times 512$ pixels.
The optimization is performed through ADAM algorithm with learning rate $0.001$.
At each iteration, we perturb the \gls{cnn} input noise $\z$ with additive white gaussian noise with standard deviation $0.1$ to strengthen the convergence. This way, the network is more robust with respect to the specific noise realization and it is forced to learn higher level features.
% the required GPU memory is 8GB. 
Without any specific code optimization, we reach a computation speed of $5$ iterations per second on a Nvidia Tesla V100 GPU, requiring $8$ GB of GPU memory.

% parametri di assemblaggio:
Concerning the proposed local post-processing and assembly in \cref{subsec:assembly}, notice that the amount of $M$ available images at the output of \gls{dip} process changes accordingly to the input image\red{, and to the PSNR requirement. Typical values are between 500 and 2500 images. The computational impact of the post-processing is mainly due to the Wiener filter that estimates the noise residual on all the $M$ images.}.
It is worth noticing that the vast majority of images needs few iterations (less than $3000$ iterations over $10000$ possible cycles, on average) to achieve the threshold of $39\textrm{dB}$ chosen to stop the inversion.
We consider multiple parameter configurations in order to include a sufficiently wide pool of investigation cases.
The block size $B$ can be chosen among $B = [32, 64, 128, 256, 512]$ pixels, and the maximum number of averaged blocks can vary as well, being $L = [1, 5, 10, 25, 50, 75, 100]$. Notice that the case $B = 512$ corresponds to select the full image, without dividing it into blocks. It is worth noticing that the configuration $\{L = 1, B = 512\}$ coincides with the absence of the local area post-processing proposed in \cref{subsec:assembly}. 

\subsection{Evaluation Metrics}

After the generation of the anonymized image $\hat{\I}$, we compute the \gls{ncc} between $\hat{\I}$ and the source device \gls{prnu} $\K$, together with the \gls{psnr} between $\hat{\I}$ and the original image $\I$. These values are the used metrics for evaluating the results and comparing with state-of-the-art.
The lower the achieved \gls{ncc} together with a high \gls{psnr}, the better the image anonymization performance.

To summarize results related to the achieved \glspl{ncc}, we make use of \gls{roc} curves related to the source device identification problem. 
Given a fixed device PRNU $\K$, \glspl{ncc} of anonymized images taken with that device are defined as the positive set, while \glspl{ncc} of images shot by other cameras are the negative set. Anonymization performance is evaluated through the \gls{auc}, as done in \cite{Entrieri2016, Mandelli2017, Bonettini2018}. Our goal is to reduce the \gls{auc} of the curves, thus making the PRNU-based identification not working, at the same time maintaining high values of \gls{psnr}.

\section{Results and Discussion}
\label{sec:results}

In this section we provide the numerical results achieved with our experimental campaign that demonstrate the capability and limitations of our methodology.
First, we deploy our method when the reference PRNU of the device is available at the analyst (i.e., $\P=\K)$.
Then, we show that our method can also be applied in the case of blind anonymization when the PRNU $\K$ is unknown (i.e., $\P=\W)$.
Results are compared with state-of-the-art techniques to highlight pros and cons.

% In this paper it is shown that simple operations do not work: \url{https://link.springer.com/content/pdf/10.1007/s11042-020-09133-9.pdf}.
% This work is from ICIP2015, it seems easy to be implemented: \url{https://ieeexplore.ieee.org/stamp/stamp.jsp?tp=&arnumber=7351088}
% This work is from ICIP2016, well written but mega theorico papocchio: \url{https://ieeexplore.ieee.org/stamp/stamp.jsp?tp=&arnumber=7533092&tag=1}.
\subsection{PRNU-aware anonymization}

\begin{figure}[t]
	\centering	\includegraphics[width=.9\columnwidth]{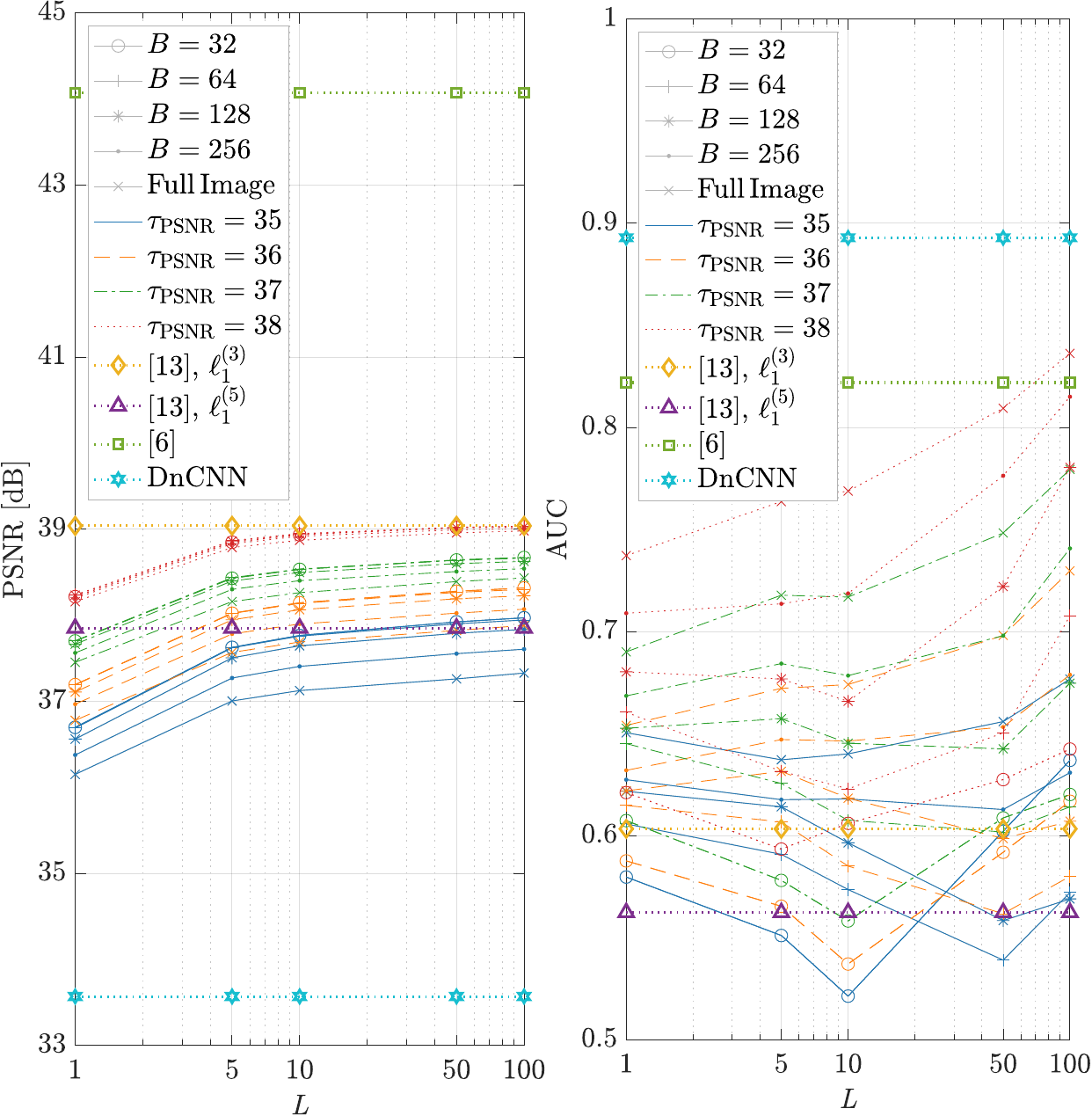}
	\caption{PRNU-aware anonymization results for the $\mathcal{D}_u$ dataset.}
	\label{fig:dresden_prnu_clean_png}
\end{figure}

\begin{figure}[t]
	\centering	\includegraphics[width=.9\columnwidth]{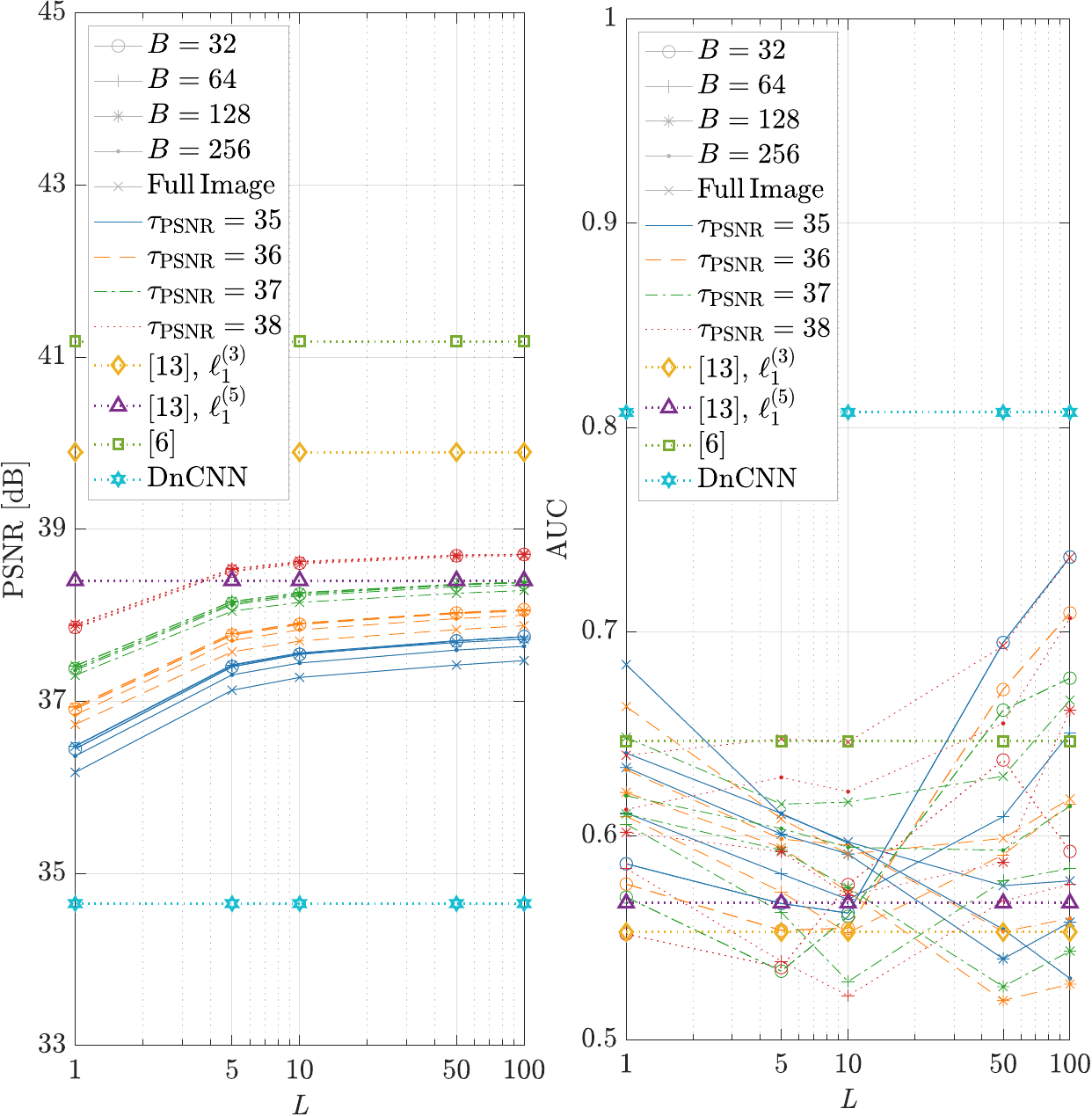}
	\caption{PRNU-aware anonymization results for the $\mathcal{D}_c$ dataset.}
	\label{fig:dresden_prnu_clean_jpg}
\end{figure}

\begin{figure}[t!]
	\centering	\includegraphics[width=.9\columnwidth]{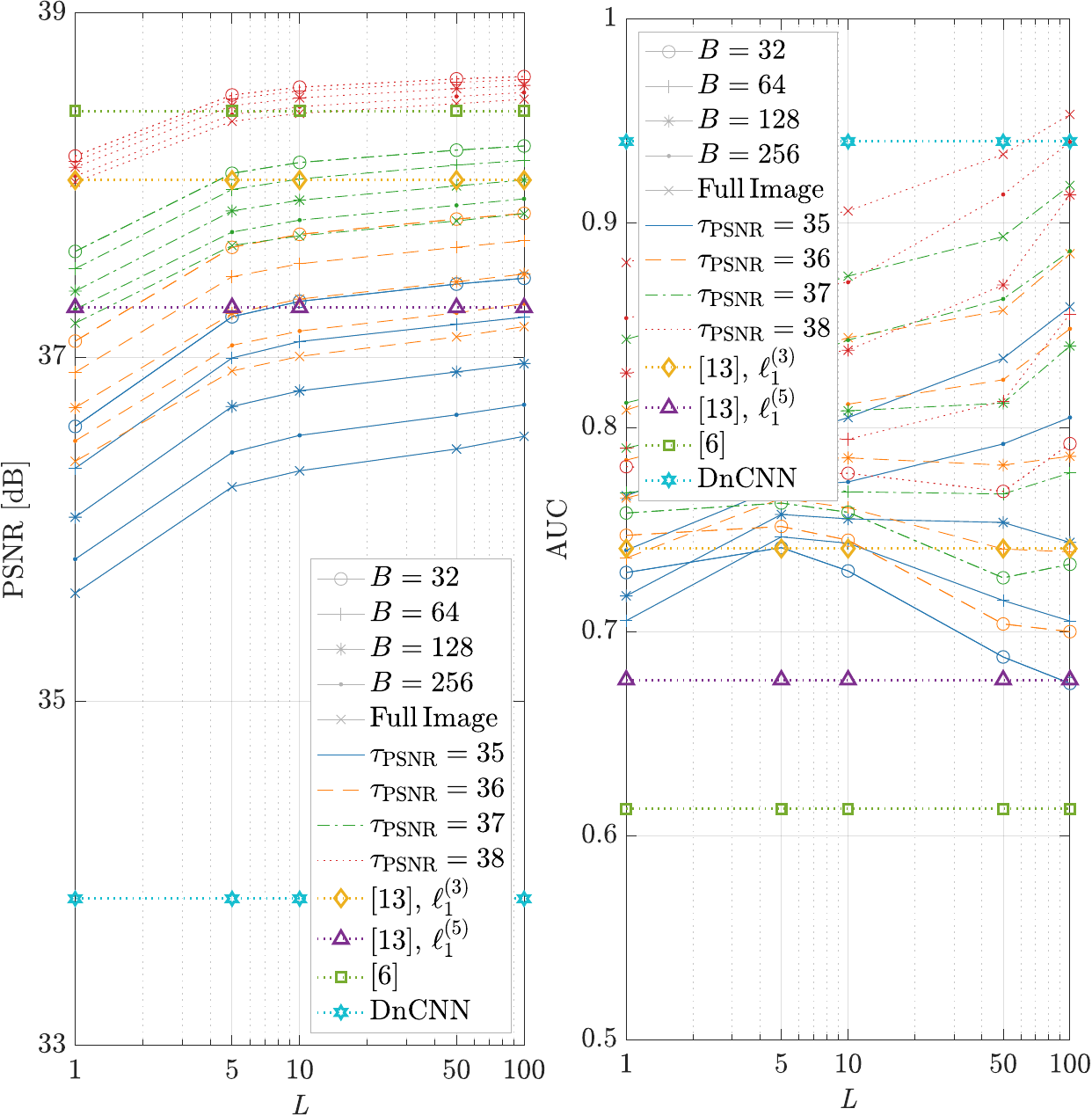}
	\caption{PRNU-aware anonymization results for the $\mathcal{V}$ dataset.}
	\label{fig:vision_prnu_clean}
\end{figure}

In this scenario, the actual PRNU of the source device to be anonymized is known. 
Results are shown in terms of PSNR and AUC of the ROC curves for the three investigated datasets.
\cref{fig:dresden_prnu_clean_png}, \cref{fig:dresden_prnu_clean_jpg}, \cref{fig:vision_prnu_clean} refer to datasets $\mathcal{D}_u$, $\mathcal{D}_c$ and $\mathcal{V}$, respectively.
\red{For the sake of clarity in the following discussion, results for $L=25$ and $L=75$ are not depicted in these plots.}
It is important to notice that results must be analyzed by watching PSNR and AUC concurrently.
Indeed, high PSNR is a good result only if paired with low AUC.
We therefore privilege solutions providing a good PSNR / AUC trade-off.
To ease the readability of reported results, we separately analyze in brief paragraphs the performance of each PRNU-anonymization method.\\
\noindent\textbf{Proposed DIPPAS method. }
% AGGIUNGERE CHE: il local post-processing serve. confronto con situazione senza post-processing --> sia il psnr che l'auc ne risentono, in tutti i casi. 
% in generale, mediare piu immagini aiuta a migliorare il PSNR
For all the investigated datasets, the proposed method is able to achieve \glspl{psnr} greater than $38$ dB, provided that a sufficiently high threshold $\tpsnr$ is chosen. 
Also in terms of \glspl{auc}, the proposed method can cover a wide range of possibilities, according to the chosen block size $B$ and amount of averaged blocks $L$.
It is worth noticing that the application of local post-processing presented in \cref{subsec:assembly} allows an improvement of the results in all the three scenarios. The \gls{dip} approach alone, which corresponds to configuration $\{L = 1, B = 512\}$, often shows low \glspl{psnr} and too high \glspl{auc}. Instead, by working on smaller image blocks, both \gls{psnr} and \gls{auc} improve, even without averaging multiple blocks together (i.e., with $L = 1$).

In general, the smaller the block size $B$, the better the \gls{psnr} achieved, even though this behaviour seems to attenuate for high values of $\tpsnr$. Besides, the more the amount of averaged blocks $L$, the better the achieved \gls{psnr}.
% In terms of \glspl{auc}, the proposed method can cover a wide range of possibilities, according to the chosen block size $B$ and amount of averaged blocks $L$.
Regarding Dresden-related datasets, middle values of $L$ seem to work better for achieving good \glspl{auc}, while dataset $\mathcal{V}$ requires higher values of $L$ for lowering the \gls{auc}.
The achieved \glspl{auc} are better on Dresden-related datasets, i.e., \cref{fig:dresden_prnu_clean_png} and \cref{fig:dresden_prnu_clean_jpg}. For these two datasets, none of the state-of-the-art works outperforms the best DIPPAS results, while $\mathcal{V}$ dataset seems to be more challenging to be anonymized.\\
%Anyway, notice that we are always able to achieve better AUCs than \cite{Mandelli2017} in every dataset, with larger or comparable PSNRs. When compared to \cite{Karakuecuek2015}, we outperform this method as well for what concerns the reported AUCs except for dataset $\mathcal{V}$.
\noindent\textbf{Proposed method in \cite{Karakuecuek2015}. }
The solution provided by \cite{Karakuecuek2015} achieves the best results in terms of PSNR for all three datasets.
However, notice that the corresponding AUCs show very poor results if compared with DIPPAS and \cite{Mandelli2017} for Dresden-related datasets. 
Concerning the dataset $\mathcal{V}$ shown in \cref{fig:vision_prnu_clean}, the AUC obtained by \cite{Karakuecuek2015} seems to outperform every proposed strategy.
\begin{figure*}[t]
	\centering	\includegraphics[width=.95\textwidth]{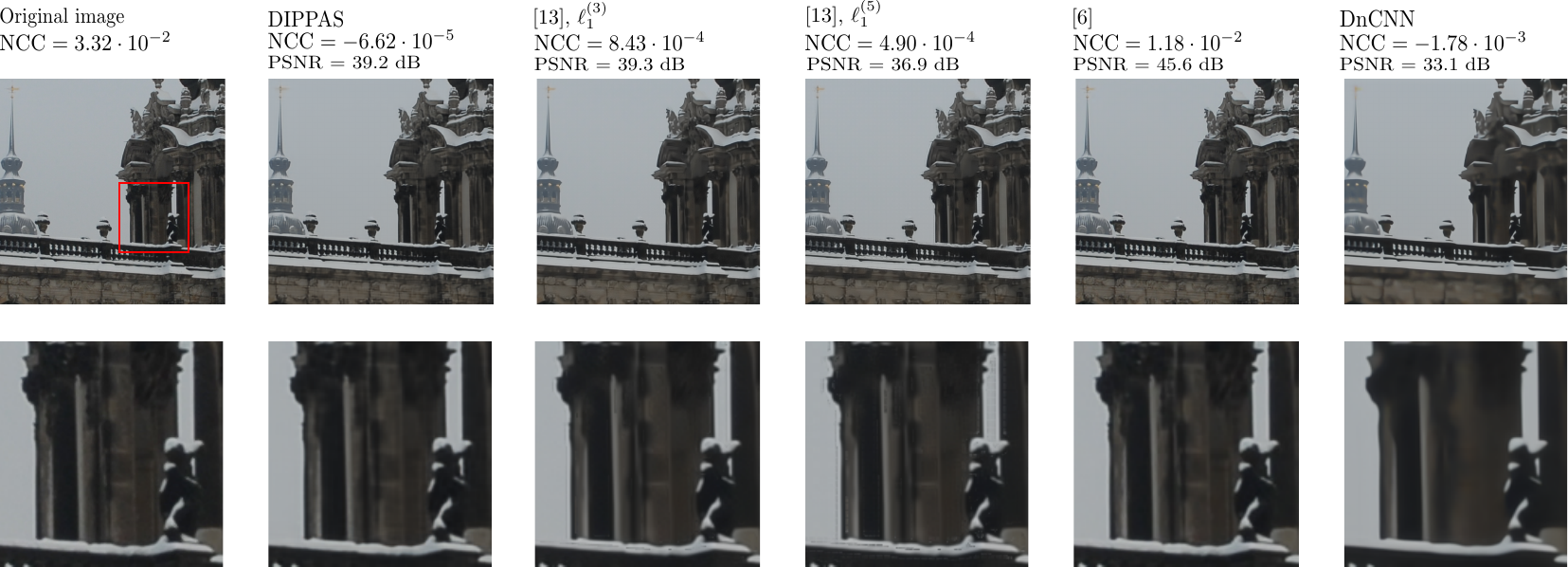}
	\caption{PRNU-aware anonymization for an image of the $\mathcal{D}_c$ dataset, comparing our proposed strategy with \cite{Mandelli2017}, \cite{Karakuecuek2015} and DnCNN. }
	\label{fig:anonymization_results}
\end{figure*}
\begin{figure*}[t]
	\centering	\includegraphics[width=.95\textwidth]{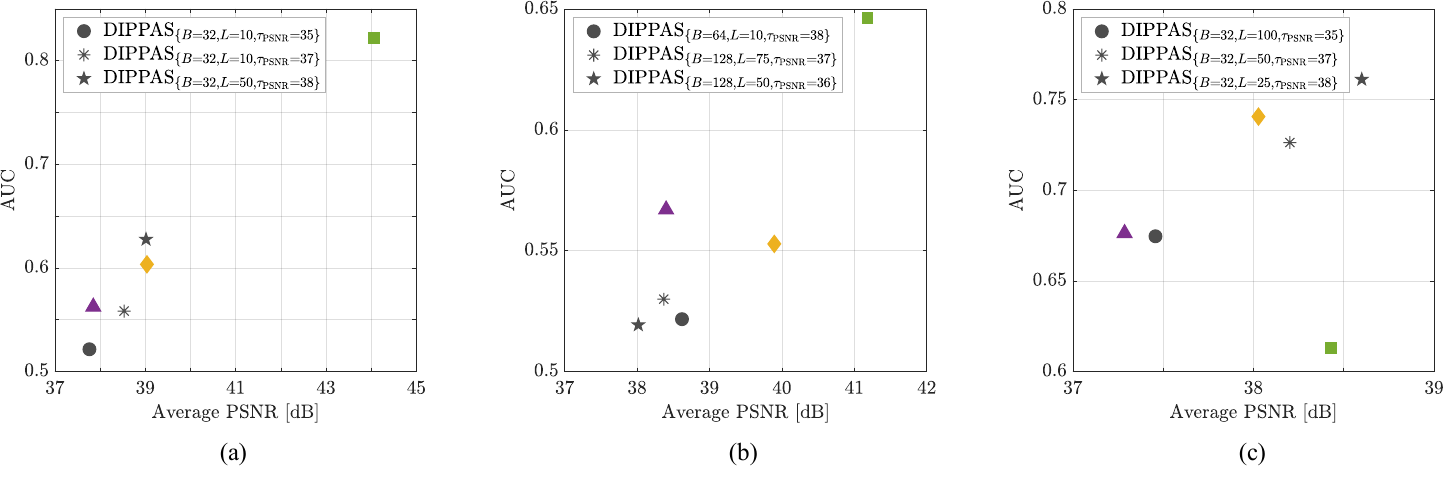}
	\caption{PRNU-aware anonymization results, reported in terms of AUC as a function of average PSNR achieved by three DIPPAS configurations for (a): the $\mathcal{D}_u$ dataset, (b): the $\mathcal{D}_c$ dataset, (c): the $\mathcal{V}$ dataset. We compare our proposed strategy with 
		\cite{Mandelli2017} $\ell_1^{(3)}$ 
		(yellow
		% 	 $\blacklozenge$, \cite{Mandelli2017} $\ell_1^{(5)}$ (purple $\blacktriangle$) and \cite{Karakuecuek2015} (green $\blacksquare$). }
		\textcolor{mat_yellow}{$\blacklozenge$}), \cite{Mandelli2017} $\ell_1^{(5)}$ (purple \textcolor{mat_purple}{$\blacktriangle$}) and \cite{Karakuecuek2015} (green \textcolor{mat_green}{$\blacksquare$}). }
	\label{fig:best_results_clean}
\end{figure*}
We think this different behaviour can be explained by the diverse nature of Vision dataset with respect to Dresden.
Indeed, in dataset $\mathcal{V}$ the device PRNU is estimated directly from JPEG-compressed images, while in Dresden-based datasets the PRNU is estimated from uncompressed ones.
As a matter of fact, the PRNU estimated from JPEG-compressed images can present artifacts due to JPEG compression, which can also contribute to hinder the subtle sensor traces left on images \cite{Chen2008}.
As a consequence, anonymizing JPEG-compressed images with respect to the PRNU estimated from uncompressed data can be slightly more complicated than anonymizing JPEG images with respect to the PRNU estimated from JPEG data.
In this vein, the strategy proposed by \cite{Karakuecuek2015} seems to work in a very accurate way only if the device PRNU is estimated from JPEG-compressed images. \\
\noindent\textbf{Proposed method in \cite{Mandelli2017}. }
The proposed strategy in \cite{Mandelli2017} achieves acceptable values of PSNRs in all the considered datasets, actually comparable to those achieved by DIPPAS. The resulting AUCs show satisfying values as well, except for the Vision-related dataset, where \cite{Mandelli2017} seems to suffer more with respect to Dresden-related datasets, following a similar trend to that previously shown by DIPPAS method. 
Regardless, notice that DIPPAS can outperform the AUCs achieved by \cite{Mandelli2017} in all three datasets.\\
\noindent\textbf{Proposed method in \cite{Zhang2017}. }
The DnCNN solution proposed in \cite{Zhang2017} shows small values of PSNRs in all the experiments. 
Furthermore, the achieved results report too high AUCs, actually unacceptable for satisfactory image anonymization.
DnCNN results seem to confirm that \red{this simple} image denoiser cannot accurately delete PRNU traces \cite{bernacki2020}, leading to poor anonymization performances.

\begin{figure}[t]
	\centering	\includegraphics[width=.9\columnwidth]{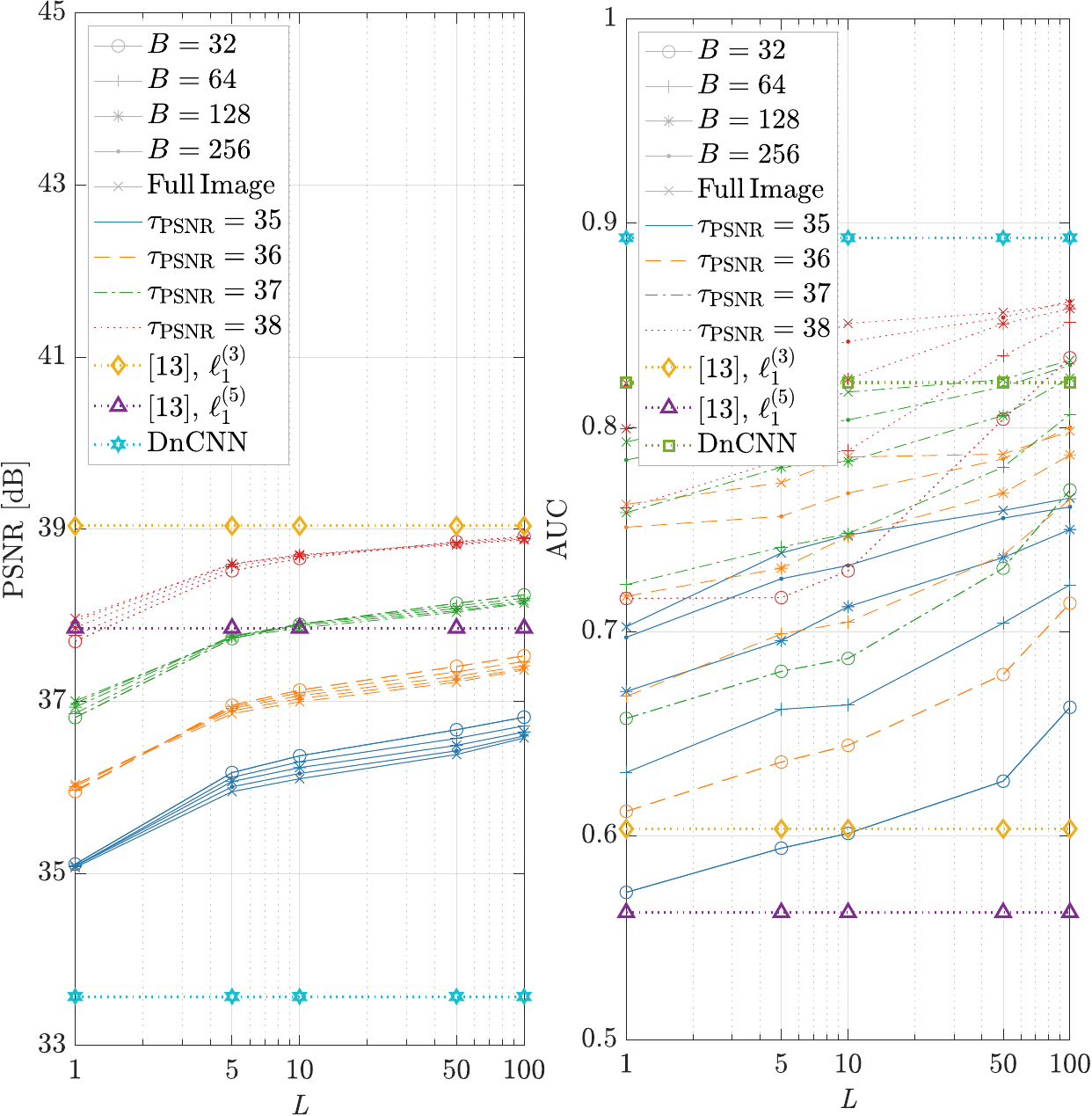}
	\caption{PRNU-blind anonymization results for the $\mathcal{D}_u$ dataset.}
	\label{fig:dresden_prnu_blind_png}
\end{figure}
\begin{figure}[t]
	\centering	\includegraphics[width=.9\columnwidth]{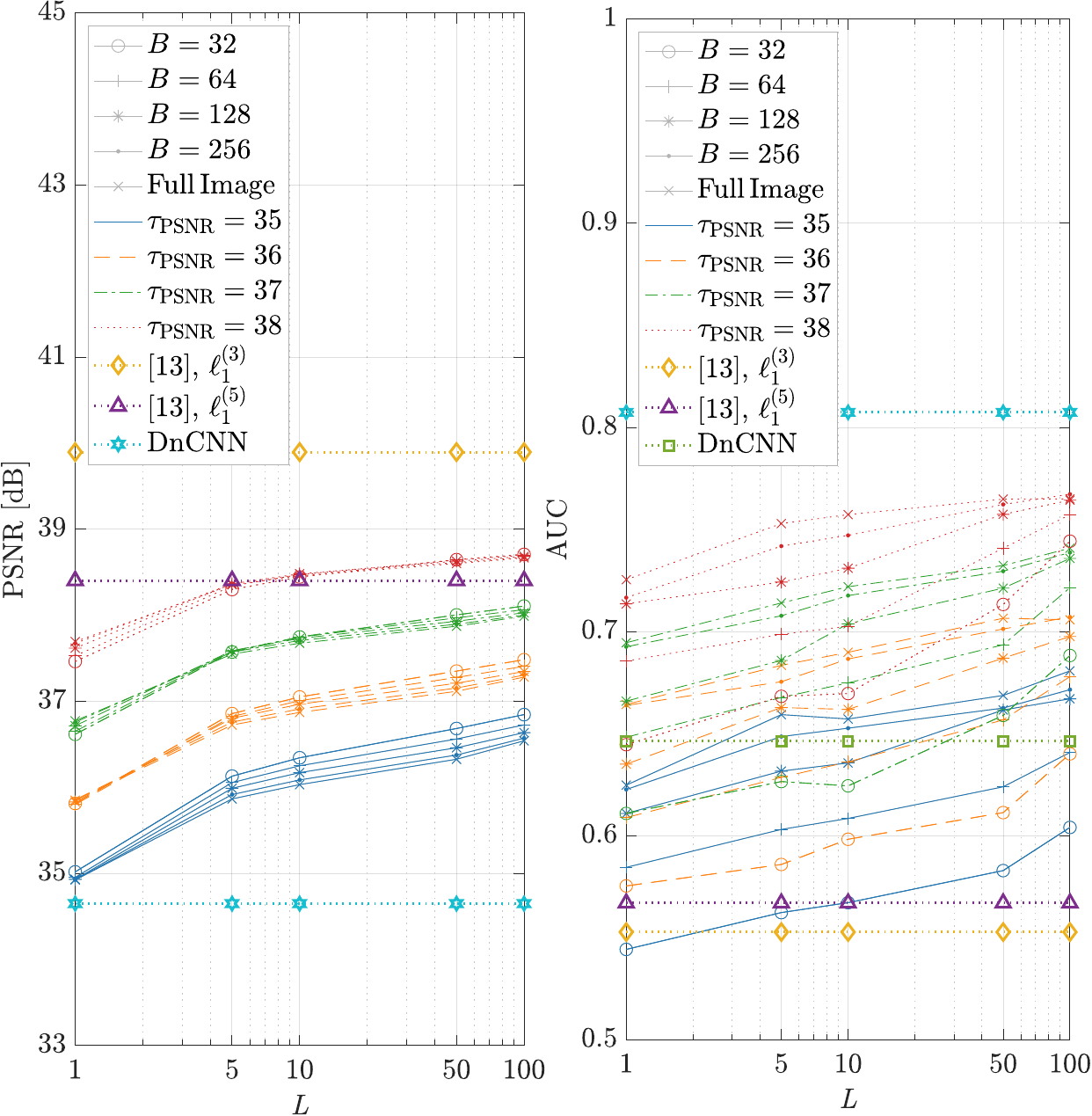}
	\caption{PRNU-blind anonymization results for the $\mathcal{D}_c$ dataset.}
	\label{fig:dresden_prnu_blind_jpg}
\end{figure}
\begin{figure}[t!]
	\centering	\includegraphics[width=.9\columnwidth]{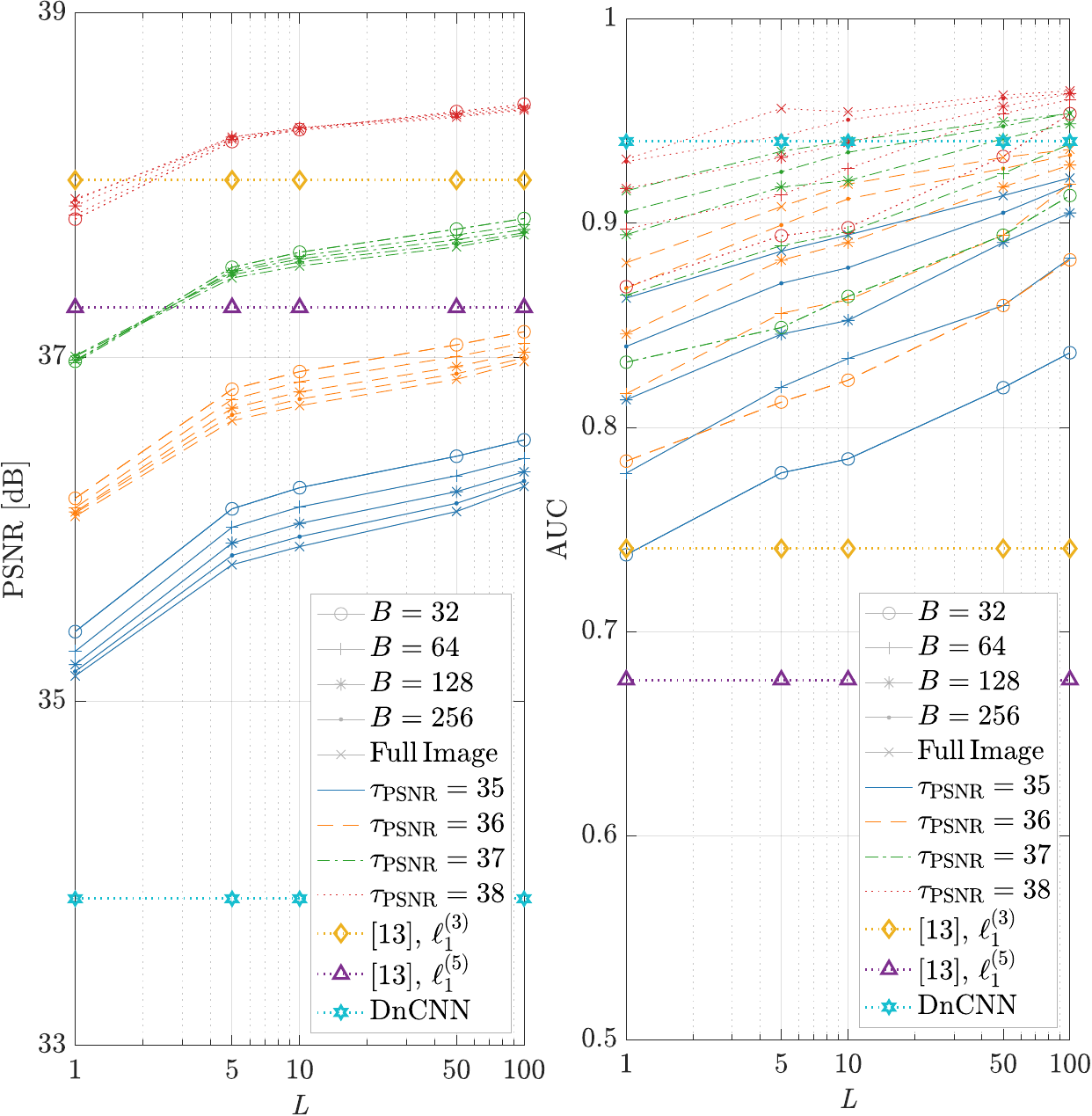}
	\caption{PRNU-blind anonymization results for the $\mathcal{V}$ dataset.}
	\label{fig:vision_prnu_blind}
\end{figure}

\cref{fig:anonymization_results} reports an example of anonymization performed over an image of $\mathcal{D}_c$ dataset. We depict the original image and its anonymized versions exploiting DIPPAS and the methods of \cite{Mandelli2017}, \cite{Karakuecuek2015} and DnCNN \cite{Zhang2017}.
%Referring at \cref{fig:dresden_prnu_clean_jpg},  
Specifically,
we choose the best performing DIPPAS parameter configuration in terms of both PSNR and AUC, i.e., $\{\tpsnr = 38, B = 64, L = 10 \}$;  we select this configuration by referring to results shown in  \cref{fig:dresden_prnu_clean_jpg}. 
Zooming in the images (red squared area), we can visually notice that the best results are obtained by DIPPAS and \cite{Karakuecuek2015}; DnCNN results in a heavily smoothed image, while \cite{Mandelli2017} introduces some edge artifacts. The method devised in \cite{Karakuecuek2015} is able to halve the original NCC, while DIPPAS dramatically scales the original NCC value by a factor of $0.002$.

To summarize the previously reported results, \cref{fig:best_results_clean} shows the behaviour of AUC as a function of the average PSNR achieved by DIPPAS in three selected parameter configurations. We compare our results with state-of-the-art as well. 
The best working condition consists in high PSNR and low AUC.
It is possible to notice that DIPPAS provides the best trade-off on Dresden dataset, and the second best one on Vision. In this latter scenario, the best trade-off is provided by \cite{Karakuecuek2015}, which however reports very poor results on Dresden dataset.
% \textcolor{sec_color}{\subsection{PRNU-blind anonymization}}

% come mai vien fuori verde??????????
\subsection{PRNU-blind anonymization}

In this scenario, the actual device PRNU is unknown, therefore the reference device fingerprint used during the \gls{dip} inversion and the blocks assembly corresponds to the noise residual extracted from the image, i.e.,  $\P = \W$.
As previously done, we report the \gls{psnr} and \gls{auc} of the \gls{roc} curves for all three investigated datasets. \cref{fig:dresden_prnu_blind_png}, \cref{fig:dresden_prnu_blind_jpg}, \cref{fig:vision_prnu_blind} depicts results for datasets $\mathcal{D}_u$, $\mathcal{D}_c$ and $\mathcal{V}$, respectively.
\red{For the sake of clarity in the following discussion, results for $L=25$ and $L=75$ are not depicted in these plots.}

In terms of \gls{psnr}, on Vision dataset we are able to outperform \cite{Mandelli2017}, while for Dresden dataset we achieve slightly lower results.
%For what concerns PSNRs, the same reasoning of previous figures.
%While for Vision dataset we are able to outperform \cite{Mandelli2017}, for Dresden-related datasets our PSNRs are slightly lower than state-of-the-art.
% Regarding the source anonymization task, 
Moreover, \gls{dippas} achieves slightly higher \glspl{auc} than state-of-the-art solutions.
%Also concerning AUCs, state-of-the-art method perform almost always sligthly better than our proposal.
Notice that the best \gls{auc} values are obtained for $L = 1$, i.e., without performing block averaging.

\iffalse
\begin{figure*}[t]
	\centering	\includegraphics[width=.85\textwidth]{figures/best_results_blind.pdf}
	\caption{PRNU-blind anonymization results, reported in terms of AUC as a function of average PSNR achieved by three DIPPAS configurations, for (a): the $\mathcal{D}_u$ dataset, (b): the $\mathcal{D}_c$ dataset, (c): the $\mathcal{V}$ dataset, comparing our proposed strategy with \cite{Mandelli2017} $\ell_1^{(3)}$ (yellow
		$\blacklozenge$) and \cite{Mandelli2017} $\ell_1^{(5)}$ (purple $\blacktriangle$).}
	%\textcolor{mat_yellow}{$\blacklozenge$}) and \cite{Mandelli2017} $\ell_1^{(5)}$ (purple \textcolor{mat_purple}{$\blacktriangle$}).}
	\label{fig:best_results_blind}
\end{figure*}
\fi
We think this less effective anonymization with respect to the previous \gls{prnu}-aware scenario can be due to the assumption done during the \gls{dip} inversion \eqref{eq:dip_prnu} in order to estimate the anonymized image. As a matter of fact, the \gls{dip} paradigm leverages the \gls{prnu}-based image modeling reported in \eqref{eq:I_prnu_def}. Whenever the \gls{prnu} estimate $\K$ is unknown and the noise residual $\W$ is used instead, as reported in \eqref{eq:dip_prnu_blind}, the model is not clearly satisfied and the \gls{dip} solution will be sub-optimum.
For this reason, in a \gls{prnu}-blind scenario, \gls{dippas} is still able to achieve good performance in terms of \gls{psnr}, but the \gls{auc} is not as good as in the \gls{prnu}-aware scenario.
%however not outperforming the method \cite{Mandelli2017} for what concerns the \glspl{auc}. 

%Following the same reasoning previously done for the \gls{prnu}-aware scenario, we show in \cref{fig:best_results_blind} the \gls{auc} behaviour as a function of the average \gls{psnr}, selecting three diverse \gls{dippas} parameter configurations for each dataset.
From these results it may seem that \gls{dippas} cannot achieve better results than \cite{Mandelli2017}, but an important point has to be noticed.
Indeed, \cite{Mandelli2017} applies different processing to edges and to flat regions, thus removing \gls{prnu} traces in concentrated local regions.
%In the next section we investigate the effect of this choice.
%
%\subsection{Comparison with \cite{Mandelli2017} along image edges}
%
%Since method \cite{Mandelli2017} outperforms our \gls{prnu}-blind proposed solution in the majority of cases, we propose a further analysis in order to carefully compare the two strategies.
Precisely, notice that method \cite{Mandelli2017} works in two separate steps: (i) estimate an anonymized version of the image exploiting inpainting techniques; (ii) substitute a denoised version of the edges extracted from the original image into the anonymized image, in order to enhance the output visual quality.
In light of these considerations, we think that the edge processing operation performed on the output image can be the weak link in the proposed pipeline of \cite{Mandelli2017}. Indeed, image edges only undergo two successive steps of BM3D denoising algorithm \cite{Dabov2007}, thus they reasonably contain enough PRNU traces for performing source attribution, as suggested in \cite{bernacki2020}.

Therefore, we compare \gls{dippas} and \cite{Mandelli2017} only along the image edges, extracted following the same pipeline proposed in \cite{Mandelli2017}.
For every dataset, we evaluate \gls{dippas} results for a parameter configuration which returns the nearest \gls{psnr} value to that achieved by \cite{Mandelli2017}. 
For instance, looking at \cref{fig:dresden_prnu_blind_png}, dataset $\mathcal{D}_u$ is evaluated for $\{ B = 32, L = 100, \tpsnr = 38 \}$ if compared to \cite{Mandelli2017}, $\ell_1^{(3)}$; we use $\{ B = 512, L = 10, \tpsnr = 37 \}$ when comparing to \cite{Mandelli2017}, $\ell_1^{(5)}$. 

We compare results in terms of relative change of \gls{auc} evaluated over image edges with respect to the \gls{auc} achieved on the full image. In a nutshell, the relative change in \gls{auc} can be computed as $(\textrm{AUC}_\text{edges} - \textrm{AUC}) / {\textrm{AUC}}$, being \gls{auc} the metrics associated to the full image. \cref{tbl:icip_edges} reports the results.
Notice that the relative \gls{auc} change maintains a coherent behaviour for all the three datasets. 
On one side, \cite{Mandelli2017} always reports a positive relative change; on the other side, \gls{dippas} presents a negative relative change. 

\gls{dippas} results are coherent with what actually happens on natural images when compared with the PRNU in a reduced region (e.g., only along the edges). Indeed, the \gls{ncc} drops as the image content is reduced, thus the \gls{auc} of the source attribution problem decreases.
On the contrary, accuracy of \cite{Mandelli2017} evaluated only along image edges strongly drops as the \gls{ncc}  increases, with a consequent \gls{auc} growth. This phenomenon can be explained by the previously reported consideration, that is, \cite{Mandelli2017} performs only denoising along the edges and this is usually not enough to hinder \gls{prnu} traces \cite{bernacki2020}. 

As a consequence, the anonymization algorithm proposed in \cite{Mandelli2017} can be easily spotted and defeated just by analyzing image edges, whereas the \gls{dippas} proposed solution does not present this drawback. Even if an analyst only use edges for \gls{prnu}-based attribution, the image would look anonymized. This is an additional pro of \gls{dippas} technique.

\begin{table}[t]
	\caption{AUC relative change on image edges, for prnu-blind anonymization. }
	\label{tbl:icip_edges}
	\resizebox{.27\columnwidth}{!}{
		\begin{minipage}{.3\columnwidth}
			\centering
			\begin{tabular}{ccc}
				NCC area& 
				\cite{Mandelli2017}, $\ell^{(3)}_1$ & 
				DIPPAS
				\\  \midrule[1.3pt]
				%\Xhline{2\arrayrulewidth}
				$\mathcal{V}_{\textrm{edges}}$ &
				$ + 11.6 \%$ &
				$- 9.1 \%$ \\  \midrule[0.1pt]
				% $\mathcal{V}_{\textrm{ w/o edges}}$  &
				% $ - 11.3 \%$
				% %\colorbox{yellow!40}{$70.34\%$}  
				% & 
				% $- 8.6 \%$ \\ \midrule[.7pt]
				$\mathcal{D}_{u_{\textrm{edges}}}$ &
				$+10.9 \%$ &
				$-17.8 \%$ \\  \midrule[0.1pt]
				% $\mathcal{D}_{u_{\textrm{ w/o edges}}}$  &
				% $-8.3\%$
				% %\colorbox{yellow!40}{$70.34\%$}  
				% & 
				% $-21.7\%$ \\ \midrule[.7pt]
				$\mathcal{D}_{c_{\textrm{edges}}}$ &
				$ + 8.7\%$ &
				$- 17.6 \%$ \\  %\midrule[0.1pt]
				% $\mathcal{D}_{c_{\textrm{ w/o edges}}}$  &
				% $+ 13.6 \%$
				% %\colorbox{yellow!40}{$70.34\%$}  
				% & 
				% $-23.5\%$ \\
			\end{tabular}
	\end{minipage}}
	\hfil
	\raggedright
	\resizebox{.25\columnwidth}{!}{
		\begin{minipage}{.3\columnwidth}
			\centering
			\begin{tabular}{ccc}
				NCC area & 
				\cite{Mandelli2017}, $\ell^{(5)}_1$& 
				DIPPAS
				\\  \midrule[1.3pt]
				%\Xhline{2\arrayrulewidth}
				$\mathcal{V}_{\textrm{edges}}$ &
				$ + 22.3 \%$ &
				$ - 9.9 \%$ \\  \midrule[0.1pt]
				% $\mathcal{V}_{\textrm{ w/o edges}}$ &
				% $ + 9.7 \%$
				% &
				% $ - 15.4 \%$ \\ \midrule[.7pt]
				$\mathcal{D}_{u_{\textrm{edges}}}$ &
				$+19.1 \%$ &
				$- 19.1 \%$ \\  \midrule[0.1pt]
				% $\mathcal{D}_{u_{\textrm{ w/o edges}}}$  &
				% $+10.2 \%$
				% %\colorbox{yellow!40}{$70.34\%$}  
				% & 
				% $-21.5\%$ \\ \midrule[.7pt]
				$\mathcal{D}_{c_{\textrm{edges}}}$ &
				$ + 6.1 \%$ &
				$- 17.7 \%$ \\  %\midrule[0.1pt]
				% $\mathcal{D}_{c_{\textrm{ w/o edges}}}$  &
				% $+ 11.3\%$
				% %\colorbox{yellow!40}{$70.34\%$}  
				% & 
				% $-22.5\%$ \\
			\end{tabular}
	\end{minipage}}
\end{table}

\section{conclusions}
\label{sec:conclusions}
\glsreset{dip}
%In this manuscript we devise a PRNU anonymization scheme for natural images that leverages the deep prior paradigm.
In this manuscript, we propose a source device anonymization scheme that leverages the \gls{dip} paradigm to attenuate \gls{prnu} traces in natural images, paired with a post-processing scheme that exploits multiple images.
%Within this method, a CNN learns to generate a PRNU-free image out of a noise realization by minimizing the distance between the input query image and the CNN output, which the PRNU has been injected into.
With this method, a \gls{cnn} learns to iteratively generate images from noise realizations. 
Specifically, at each \gls{dip} iteration: (i) the \gls{cnn} generates an image; (ii) we inject the device \gls{prnu} into this image; (iii) we minimize the distance between the input query image and the \gls{prnu}-injected image. In doing so, we are able to generate multiple images with a strongly attenuated \gls{prnu} pattern and high visual quality.
%To this purpose, we define the PRNU anonymization task as an inverse problem. Then e recast such problem as a DIP problem, finding the CNN parameters that produce the best estimate of the PRNU-free image.
Finally, 
%for the CNN optimization is agnostic of the NCC computed for device attribution, 
we devise an efficient post-processing operation for assembling the final anonymized image from the \gls{cnn} outputs realized at different iterations.

We compare our method against state-of-the-art anonymization schemes through numerical examples. In particular, when the \gls{prnu} of the device is available, we achieve our best results.
Our scheme can be generalized to the case of blind anonymization, i.e., when the device \gls{prnu} is unknown and only a noise residual can be extracted from the query image and then injected into the \gls{dip} generation process.

%However, as expected, the average performances decrease.
Not surprisingly, our method suffers when the injected noise is quite different from the source device \gls{prnu}. However, it still proves interesting when compared to state-of-the-art solutions if we consider the homogeneity of \gls{prnu} removal effect. Indeed, we are capable of removing \gls{prnu} traces on all image regions, whereas the considered baseline leaves image edges mainly non-anonymized.

Our future work will be devoted to investigate the possibility of starting from a pre-trained network to speed-up convergence. Moreover, we will focus on a better inversion model to be used in case of blind \gls{prnu} removal.

%%%%%%%%%%%%%%%%%%%%%%%%%%%%%%%%%%%%%%%%%%%%%%
%%                                          %%
%% Backmatter begins here                   %%
%%                                          %%
%%%%%%%%%%%%%%%%%%%%%%%%%%%%%%%%%%%%%%%%%%%%%%
\printglossary[nonumberlist]

\begin{backmatter}

\section*{Availability of supporting data}
All the numeric experiments shown in this manuscript relies on publicly available datasets \cite{dresden, vision}.
The codes have been released in a GitHub repository at:
\url{https://github.com/polimi-ispl/dip_prnu_anonymizer}.

\section*{Competing interests}
The authors declare that they have no competing interests.

\section*{Funding}
The authors have no funding to be declared.

\section*{Author's contributions}
All authors contributed in the design of the proposed method and writing the manuscript.

\section*{Acknowledgments}
This work was supported by the PREMIER project, funded by the Italian Ministry of Education, University, and Research within the PRIN 2017 program.

This material is based on research sponsored by the Defense Advanced Research Projects Agency (DARPA) and the Air Force Research Laboratory (AFRL) under agreement number FA8750-20-2-1004. The U.S. Government is authorized to reproduce and distribute reprints for Governmental purposes notwithstanding any copyright notation thereon. The views and conclusions contained herein are those of the authors and should not be interpreted as necessarily representing the official policies or endorsements, either expressed or implied, of DARPA and AFRL or the U.S. Government.

Hardware support was generously provided by the NVIDIA Corporation.

% if your bibliography is in bibtex format, use those commands:
\bibliographystyle{bmc-mathphys} % Style BST file (bmc-mathphys, vancouver, spbasic).
\bibliography{biblio}      % Bibliography file (usually '*.bib' )

\end{backmatter}
\end{document}